\documentclass[reprint, amsmath, amssymb, aps, superscriptaddress]{revtex4-2}
\usepackage{format}
\usepackage{soul} 
\usepackage[dvipsnames]{xcolor}
\usepackage{verbatim}
\allowdisplaybreaks

\usepackage{soul} 
\usepackage{ulem}
\usepackage{hyperref}
\usepackage{orcidlink}
\begin{document}

\preprint{APS/123-QED}

\title{ A Protocol for Shielding-Enhanced Loading of Single Polar Molecules \\ into Optical Tweezers}

\author{Reuben R. W. Wang\orcidlink{0000-0002-1069-9746}} \thanks{These authors contributed equally to this work.}
\affiliation{ ITAMP, Center for Astrophysics $|$ Harvard \& Smithsonian, Cambridge, Massachusetts 02138, USA }
\affiliation{ Department of Physics, Harvard University, Cambridge, Massachusetts 02138, USA }

\author{Christian H. Nunez\orcidlink{0009-0007-4886-6195}} \thanks{These authors contributed equally to this work.}
\author{Conner Williams\orcidlink{0000-0003-2138-6777}}
\author{Amanda Younes\orcidlink{0009-0004-1985-6113}}
\author{Li Du\orcidlink{0000-0002-5940-0143}}
\affiliation{ Department of Physics, Harvard University, Cambridge, Massachusetts 02138, USA }
\affiliation{ Department of Chemistry and Chemical Biology, Harvard University, Cambridge, Massachusetts 02138, USA }
\affiliation{ Harvard-MIT Center for Ultracold Atoms, Cambridge, Massachusetts 02138, USA }

\author{H. R. Sadeghpour\orcidlink{0000-0001-5707-8675}}
\affiliation{ ITAMP, Center for Astrophysics $|$ Harvard \& Smithsonian, Cambridge, Massachusetts 02138, USA }

\author{Kang-Kuen Ni\orcidlink{0000-0002-0537-0719}}
\affiliation{ Department of Physics, Harvard University, Cambridge, Massachusetts 02138, USA }
\affiliation{ Department of Chemistry and Chemical Biology, Harvard University, Cambridge, Massachusetts 02138, USA }
\affiliation{ Harvard-MIT Center for Ultracold Atoms, Cambridge, Massachusetts 02138, USA }

\date{\today} 

\begin{abstract}

We propose the high-fidelity preparation of single bosonic molecules in optical tweezers starting from small tweezer-trapped molecular ensembles. Our scheme combines a static electric field and a microwave field to generate strong, tunable, anisotropic interactions that shield the molecules against two-body collisional loss. We show that this shielding eliminates all long-range bound states, preventing three-body recombination. This elimination persists for all microwave ellipticities, including the experimentally practical limit of linear polarization. Application of an additional electric field gradient can be used to induce controlled spilling of strongly interacting molecules out of the trap until one remains. With realistic experimental parameters, we estimate that single tweezer-trapped NaCs molecules can be isolated from a pair with fidelities exceeding 99\%, and $> 95\%$ per site across an array. These results establish collisional shielding with electric fields as an effective tool for preparing highly-filled tweezer arrays of polar molecules.

\end{abstract}

\maketitle

\section{ Introduction \label{sec:introduction} } 

Ultracold polar molecules offer strong long-range interactions and long-lived quantum states that can be manipulated for a broad variety of quantum science applications~\cite{Yan13_Nat, Ni18_ChemicalScience, YaoSpinLiquid2018, Homeier_tjModel_PRL_2024, DurhamQudits_NJP_2020, VictorCoveyPreskill_MolecularQudit_PRX_2020, Fleig_FrAg_Theory2023, Carroll25_Sci}. 
Individual trapping and addressing of molecules in optical tweezer arrays~\cite{Liu18_Sci, Anderegg19_Sci, Cairncross21_PRL, Zhang22_QST, Holland23_Sci, Vilas24_Nat} has enabled coherent spin-exchange dynamics with controlled dipolar interactions between molecules, entanglement of molecules, and quantum gates between molecular qubits~\cite{Holland23_Sci, Bao23_Sci, Picard25_Nat, Ruttley25_Nat, Lu26_arxiv}. 
While molecular tweezer platforms have already demonstrated their value for quantum science, state-of-the-art experiments typically load ground state molecules in only 20--50\% of array sites~\cite{Zhang22_QST, Bao23_Sci, Holland23_Sci}. To increase the effective filling fraction, techniques to identify sites in which a ground state molecule is not present have been developed and implemented~\cite{Ruttley24_PRXQ, Picard24_PRXQ, Holland25_PRX}. However, scaling to large and highly filled tweezer arrays of molecules remains an outstanding challenge.

At the same time, tremendous progress has been made in stabilizing bulk gases of ultracold dipolar molecules.  
Key to this success are ``collisional shielding" \cite{Micheli07_PRA, Gorshkov08_PRL} techniques that employ microwave (ac) \cite{Cooper09_PRL, Karman18_PRL, Lassabliere18_PRL, Karman25_PRXQ} and static electric (dc)
\cite{Avdeenkov06_PRA, Quemener16_PRA, Wang15_NJP, GonzalezMartinez17_PRA} fields to suppress two- \cite{Matsuda20_Sci, Li21_Nat, Anderegg21_Sci, Lin23_PRX} and three-body \cite{Yuan25_arxiv} molecular losses. 
Typically, these schemes use dressed molecular rotational states and dipole-dipole coupling to create a long-range repulsive barrier that suppresses short-range inelastic processes.     
In practice, shielding has been essential for enabling evaporative cooling of dipolar molecules, and in recent years, has been used to realize molecular Bose-Einstein condensates \cite{Bigagli24_Nat, Shi26_NatPhys} and degenerate Fermi gases \cite{Schindewolf22_Nat, Lin26_arxiv}.

\begin{figure}[ht]
    \centering
    \includegraphics[width=1\linewidth]{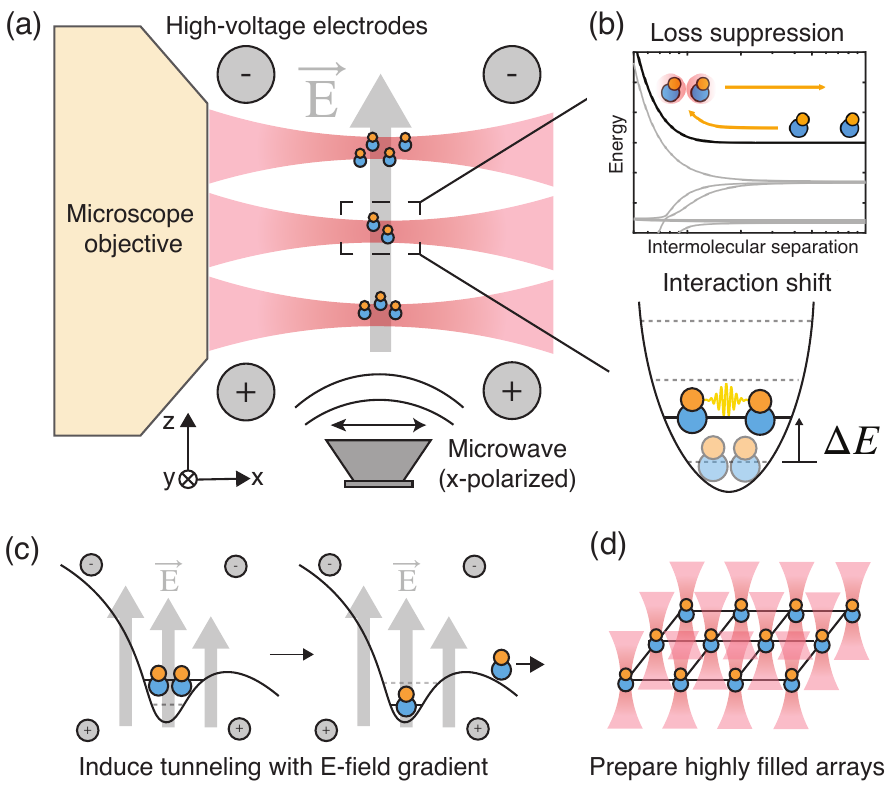}
    \vspace{-15pt}
    \caption{
    Enhanced tweezer array loading of NaCs molecules with ac-dc(X) shielding and electric field control. (a) A high-NA microscope objective projects optical tweezers containing multiple ground state molecules. High-voltage electrodes and a microwave source provide external field control for collisional shielding. (b) Collisional shielding enforces repulsive interactions between molecules, (1) suppressing two- and three-body loss and (2) creating interaction shifts in doubly-occupied tweezers. (c) An electric field gradient is applied to the polarized molecules, modifying the trap potential and inducing tunneling for molecules near the barrier peak. (d) Highly filled arrays can be prepared with this protocol. }
    \vspace{-20pt}
    \label{fig:experimental_setup}
\end{figure}

In this work, we address the challenge of scalable molecular tweezer array assembly by starting with molecular ensembles in each tweezer site. Our task is then to isolate one molecule from each ensemble. To this end, we propose a collisional shielding scheme that stabilizes co-trapped molecules and enables their controlled spilling from the trap.

Our shielding scheme uses a combination of ac and dc electric fields to suppress two-body loss and remove field-linked (FL) bound states \cite{Avdeenkov03_PRL} that would otherwise induce three-body recombination \cite{Stevenson24_PRL}. We demonstrate that this collisional shielding is bound state free for arbitrary microwave polarization ellipticity, including the practical limit of a linearly polarized microwave. The dc field in this shielding scheme serves a second purpose: to polarize the molecules in the laboratory frame, enabling external field gradients to modify the trapping potential seen by the molecules. In this way, this field configuration simultaneously stabilizes tweezer-trapped molecules against loss while facilitating controlled tunneling of molecules out of the tweezer until one remains.

The remainder of this paper is organized as follows: Sec.~\ref{sec:tweezer_loading} introduces our scheme for efficient unit-loading of tweezer trapped molecules. The required field dressing for collisional shielding is provided in Sec.~\ref{sec:shielding_scheme}, where we demonstrate that this shielding scheme removes FL bound states for any microwave polarization ellipticity, and we compute the two-body loss rates between shielded molecules. Sec.~\ref{sec:results} provides estimates for unit-loading fidelities and necessary technical requirements for realistic tweezer array systems. Finally,
Sec.~\ref{sec:conclusions} presents our conclusions and outlook.

\section{ Shielding-enhanced loading \label{sec:tweezer_loading} }

Static electric fields have been used to realize resonant collisional shielding of KRb molecules \cite{Matsuda20_Sci, Li21_Nat, Lin26_arxiv}, while the application of microwaves has been shown to provide a powerful means of control over molecular interactions \cite{Gorshkov08_PRL, Yan20_PRL, Schindewolf26_arxiv, Wang26_arxiv, Ho26_PRR}. 
Separately, utilizing magnetic field gradients to ``spill" atoms out of optical traps for preparation of few-fermion systems has been demonstrated with high fidelity in a single tweezer \cite{Serwane11_Sci}, and across a tweezer array \cite{Jain2026}. 
Inspired by these demonstrations, the objective of this work is to leverage these techniques for efficient preparation of molecular tweezer arrays, whose key components are illustrated in Fig.~\ref{fig:experimental_setup}.

We envision our protocol to commence with optical tweezers, formed with a high-NA (numerical aperture) microscope objective, that initially contain a small, stochastic number ($\lesssim 5$) of polar molecules. 
One possible pathway to these small molecular ensembles is to first create low entropy mixtures of the constituent atoms in each tweezer site and then to form ground state molecules by Feshbach association and stimulated Raman adiabatic passage directly in the tweezer.
This method of association has been demonstrated for forming single tweezer trapped ground state molecules from single pairs of atoms \cite{Cairncross21_PRL}.
Preparation of tweezers containing multiple molecules will be a prerequisite step to this scheme.

In this protocol, association of atoms into molecules occurs in the presence of a strong dc electric field \cite{Matsuda20_Sci}, applied with high-voltage electrodes, that polarizes the molecules in the lab frame and shields them against inelastic loss upon creation [Fig.~\ref{fig:experimental_setup}\textcolor{blue}{(a)}]. 
Once the molecules are formed in their appropriate quantum state, an additional $x$-polarized microwave field is applied that tunes the intermolecular forces to prevent three-body recombination and further suppress two-body loss [Fig.~\ref{fig:experimental_setup}\textcolor{blue}{(b)} upper panel, see Sec.~\ref{sec:shielding_scheme} for further details].

With lab-frame-polarized molecules, external electric field gradients can modify the tweezer potential through the differential dc Stark shift.  
In particular, a linear field gradient spills the molecules from the trap, akin to magnetic field spilling for trapped fermionic atoms~\cite{Serwane11_Sci}. 
Having eliminated three-body recombination from FL states, the molecular dynamics during spilling with $N \geq 2$ molecules is dominated by pairwise shielded interactions that enhance the tunneling rate of molecules out of the trap. 

The final step of single molecule isolation relies on the large difference in quantum tunneling timescales resultant from the two-molecule interactions.
The dressing that enables shielding and trap shaping induces an interaction shift $\Delta E$ of the doubly-occupied tweezer ground state [Fig.~\ref{fig:experimental_setup}\textcolor{blue}{(b)} lower panel]. 
This interaction increases the energy available for the first tunneling event from the doubly-occupied tweezer, resulting in rapid single-molecule tunneling out of the trap when an appropriate field gradient is applied [Fig.~\ref{fig:experimental_setup}\textcolor{blue}{(c)}]. After the first molecule tunnels out, the interaction shift is no longer present, and thus the remaining single-molecule tunneling rate is suppressed.
This separation of tunneling timescales, paired with the application of a field gradient for an optimized duration, can lead to high-fidelity unit-loading in an optical tweezer array [Fig.~\ref{fig:experimental_setup}\textcolor{blue}{(d)}, see Sec.~\ref{sec:results} for further details].

\section{ AC-DC(X) shielding 
\label{sec:shielding_scheme} } 

\subsection{ Shielding model }

\begin{figure}[ht]
    \centering
    \includegraphics[width=\linewidth]{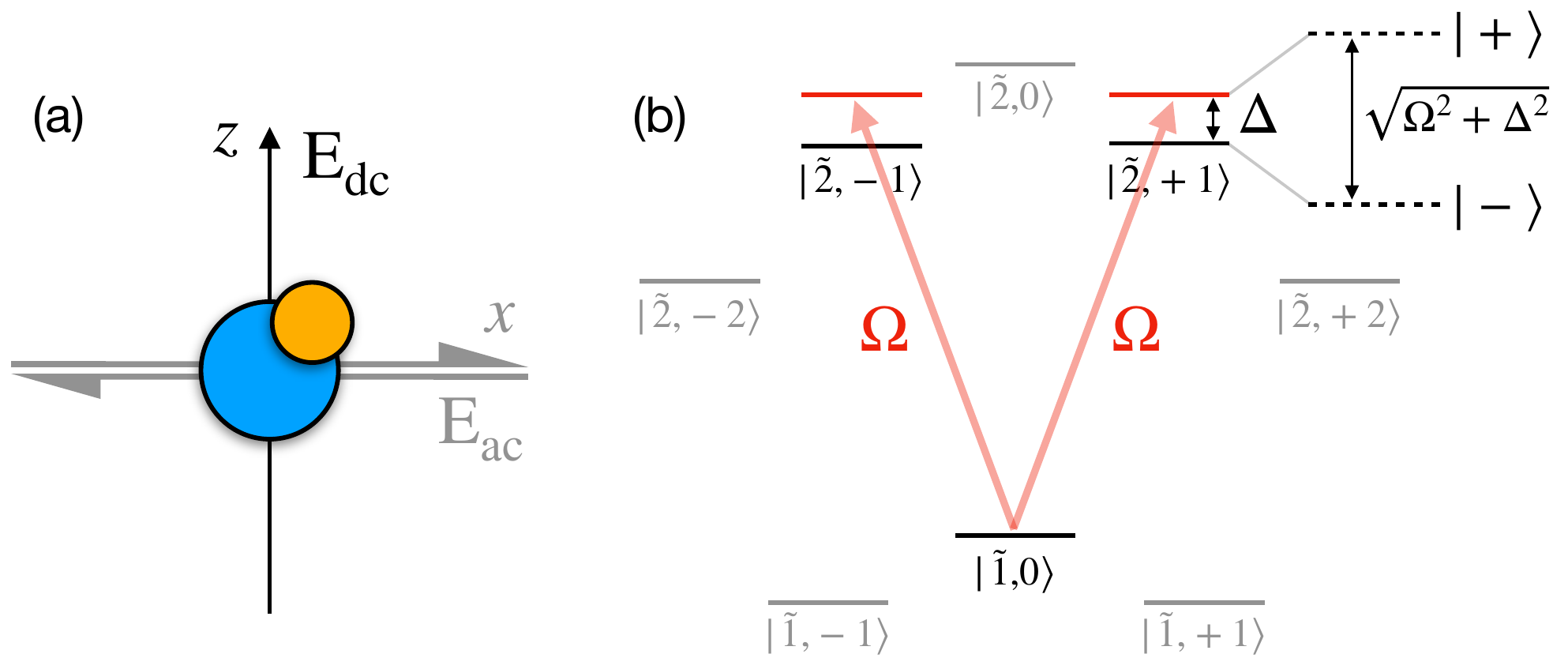}
    \caption{ The ac-dc(X) dressing scheme. (a) Depiction of a polar molecule subject to a dc electric field along $z$ and an ac microwave field polarized along $x$. (b) Illustration of the single-molecule energy level diagram subject to ac-dc dressing. }
    \label{fig:shielding_scheme_cartoon}
\end{figure}

Our shielding scheme uses a dc field to shift the rotational states into a F\"orster resonance to suppress two-body loss and an ac field to further remove FL bound states \cite{Avdeenkov03_PRL} in the long-range potential. Unlike previous proposals \cite{Wang26_arxiv, Ho26_PRR}, we show that this scheme is effective for any microwave ellipticity. 
In particular, we focus on the case where the ellipticity results in linear polarization along $\hat{\boldsymbol{x}}$, orthogonal to the applied dc electric field ($\hat{\boldsymbol{z}}$), a scheme we refer to as ac-dc(X) shielding, see Fig.~\ref{fig:shielding_scheme_cartoon}(a). 
This choice is motivated by its practical implementation in experimental setups with rod-shaped electrodes;
rods can act as polarizers to external microwave radiation, preventing easy application of circularly polarized microwaves.

A single $^{1}\Sigma$ bialkali molecule subject to ac-dc fields is well described by the dressed rigid rotor Hamiltonian,
\begin{align}
    {\cal H}
    &=
    {\cal H}_{\rm rot} 
    +
    {\cal H}_{\rm dc}
    +
    {\cal H}_{\omega}
    +
    {\cal H}_{\rm ac} \nonumber\\
    &=
    B_0 \boldsymbol{N}^2
    -
    \boldsymbol{d} \cdot \boldsymbol{{\rm E}}_{\rm dc}
    +
    \hbar\omega
    \left(
    a_{\sigma}^{\dagger} a_{\sigma} - n_{0}
    \right) \nonumber\\
    &\quad 
    -
    \frac{ {\rm E}_{\rm ac} }{ 2 \sqrt{ n_{0} } }
    \big[
    ( d_{-}\cos\xi - d_{+} \sin\xi )
    a_{\sigma}^{\dagger} 
    +
    {\rm h.c.} 
    \big], 
\end{align}
where $B_0$ is the rotational constant, $d$ is the molecular frame dipole moment, ${\rm E}_{\rm dc}$ and ${\rm E}_{\rm ac}$ are the strengths of the dc and ac fields respectively, $n_{0}$ is the background number of microwave photons, $a_{\sigma}^{\dagger}$ ($a_{\sigma}$) is the microwave creation (annihilation) operator with polarization $\sigma$, and $d_{\pm} = \mp (d_x \pm i d_y) / \sqrt{2}$ are the associated dipole operators. 
Our shielding scheme is generally formulated for any arbitrary microwave polarization, controlled by the ellipticity angle $\xi$ \cite{Karman19_PRA, Karman20_PRA}.

Considering only the combined rotational-photon states $| \tilde{N}, m \rangle\ket{ n } = \ket{ \tilde{1}, 0 }\ket{ 0 }, \ket{ \tilde{2}, -1 }\ket{ -1 }, \ket{ \tilde{2}, +1 }\ket{ -1 }$ that are closely degenerate and bright to the microwaves, the Hamiltonian has the representation
\begin{align}
    \boldsymbol{{\cal H}}
    &=
    \hbar
    \begin{pmatrix}
        0 & -\frac{ \Omega }{ 2 } \sin\xi & \frac{ \Omega }{ 2 } \cos\xi \\
        -\frac{ \Omega }{ 2 } \sin\xi & -\Delta & 0 \\
        \frac{ \Omega }{ 2 } \cos\xi & 0 & -\Delta 
    \end{pmatrix},
\end{align} 
with Rabi frequency $\Omega$ and detuning $\Delta$. 
Diagonalizing this Hamiltonian gives the dressed spectrum
\begin{subequations}
\begin{align}
    \varepsilon_{\emptyset} 
    &=
    -\hbar\Delta : \\
    &\quad
    \ket{ \emptyset }
    =
    \left(
    \sin\xi\ket{ \tilde{2}, +1 } 
    +
    \cos\xi\ket{ \tilde{2}, -1 }
    \right)
    \ket{ -1 }, 
    \nonumber\\
    \varepsilon_{\pm} 
    &=
    -\frac{ \hbar\Delta }{ 2 } \pm \frac{ \hbar }{ 2 } \sqrt{\Delta^2 + \Omega^2} : 
    \\
    &\quad 
    \ket{ \pm }
    =
    \pm c_{\mp} 
    \ket{ \tilde{1}, 0 }
    \ket{ 0 } \nonumber\\
    &\quad\quad\quad\:\:
    +
    c_{\pm} 
    \left(
    \cos\xi \ket{ \tilde{2}, +1 } 
    -
    \sin\xi \ket{ \tilde{2}, -1 }
    \right)
    \ket{ -1 }, \nonumber
\end{align}
\end{subequations}
where $c_{\pm} = [ 2 + 2 \delta( \delta \pm \sqrt{ 1 + \delta^2 } ) ]^{-1/2}$ and $\delta = \Delta/\Omega$.
Unless otherwise specified, we will set $\xi = 45^{\circ}$ ($x$-polarization), as illustrated in Fig.~\ref{fig:shielding_scheme_cartoon}. 

\begin{table}[ht]
    \centering
    \begin{tabular}{l|l|c}
        state label & $|{ \tilde{N}_A, m_{A}; \tilde{N}_B, m_{B} }\rangle_S\ket{ n }$ & energy \\
        \hline
        $|g; g\rangle\ket{0}$ & $|{ \tilde{1},0; \tilde{1},0 }\rangle_S\ket{ 0 }$ & $0$ \\
        $|F\rangle\ket{0}$ & $|{ \tilde{2}, 0; \tilde{0}, 0 }\rangle_S\ket{ 0 }$ & $-\hbar\Delta_{F}$  \\
        $|\bar{F}\rangle\ket{0}$ & $|{ \tilde{2}, + 1; \tilde{0}, 0 }\rangle_S\ket{ 0 }$ & $-\hbar\Delta_{\bar{F}}$ \\
        $|g; \bar{e}\rangle\ket{-1}$ & $|{ \tilde{1}, 0; \tilde{2}, -1}\rangle_S\ket{ -1 }$ & $-\hbar\Delta$ \\
        $|g; e\rangle\ket{-1}$ & $|{ \tilde{1}, 0; \tilde{2}, +1 }\rangle_S\ket{ -1 }$ & $-\hbar\Delta$ \\
        $|e; \bar{e}\rangle\ket{-2}$ & $|{ \tilde{2}, +1; \tilde{2}, -1 }\rangle_S\ket{ -2 }$ & $-2 \hbar\Delta$ \\
        $|e; e\rangle\ket{-2}$ & $|{ \tilde{2}, +1; \tilde{2}, +1 }\rangle_S\ket{ -2 }$ & $-2 \hbar\Delta$ \\
        $|\bar{e}; \bar{e}\rangle\ket{-2}$ & $|{ \tilde{2}, -1; \tilde{2}, -1 }\rangle_S\ket{ -2 }$ & $-2 \hbar\Delta$ 
    \end{tabular}
    \caption{ Pair molecular states that are utilized in deriving the effective shielding potentials, along with the energies relative to the incident threshold. 
    The subscript $S$ indicates that the states are symmetric under the exchange of molecules $A$ and $B$.  }
    \label{tab:basis_set}
\end{table}

The potential energy surface (PES) describing the interaction between two approaching molecules is well represented by an effective potential. This potential is derived considering only the pair molecular states that are energetically in proximity and relevant to dipolar interactions, listed in Tab.~\ref{tab:basis_set}. 
In this basis, the first-order dipole-dipole interaction in relative coordinates $\boldsymbol{r} = (r, \theta, \phi)$ is derived to be
\begin{align} \label{eq:firstorder_interaction}
    V_{\rm dd}^{(1)}(\boldsymbol{r})
    &=
    \frac{ d_{\parallel}^2 }{ 4 \pi \epsilon_0 }
    \frac{ 1 - 3 \cos^2\theta }{ r^3 } \nonumber\\
    &\quad 
    -
    \frac{ d_{\perp}^2 }{ 4 \pi \epsilon_0 }
    \frac{ 3 \sin^2\theta \cos(2\phi) }{ r^3 }
    \sin(2\xi), 
\end{align}
in terms of the effective squared dipole moments
\begin{subequations}
\begin{align}
    d_{\parallel}^2 
    &=
    ( \alpha_{-} d_{e} - \alpha_{+} d_g )^2 - \alpha_{-} \alpha_{+} d_{g\rightarrow{e}} d_{{e}\rightarrow{g}}, \\
    d_{\perp}^2
    &=
    \alpha_{-} \alpha_{+}
    d_{g \rightarrow e} d_{\bar{e} \rightarrow g},
\end{align}
\end{subequations}
where $\alpha_{\pm} = \{ [ 1 + 2 \delta \left(\delta \pm \sqrt{\delta^2 + 1}\right) ] / ( 4\delta^2 + 4 ) \}^{1/2}$ are mixing coefficients and $\delta = \Delta/\Omega$ is the relative detuning. 
The effective squared dipole moments can take negative values as they are assigned to a particular form of the anisotropic interaction potential by a choice of convention. 
The potential in Eq.~\ref{eq:firstorder_interaction} is obtained having taken that $d_{e} = d_{\bar{e}}$ and $d_{e \rightarrow g} d_{g \rightarrow e} = d_{\bar{e} \rightarrow g} d_{g \rightarrow \bar{e}}$, the induced and transition dipole moments with labels $g, e$ and $\bar{e}$ corresponding to the dc dressed rotational states tabulated in Tab.~\ref{tab:basis_set}. Explicit definitions of all dipole moments are provided in App.~\ref{app:effective_potential_derivation}.

\begin{figure}[ht]
    \centering 
    \includegraphics[width=\linewidth]{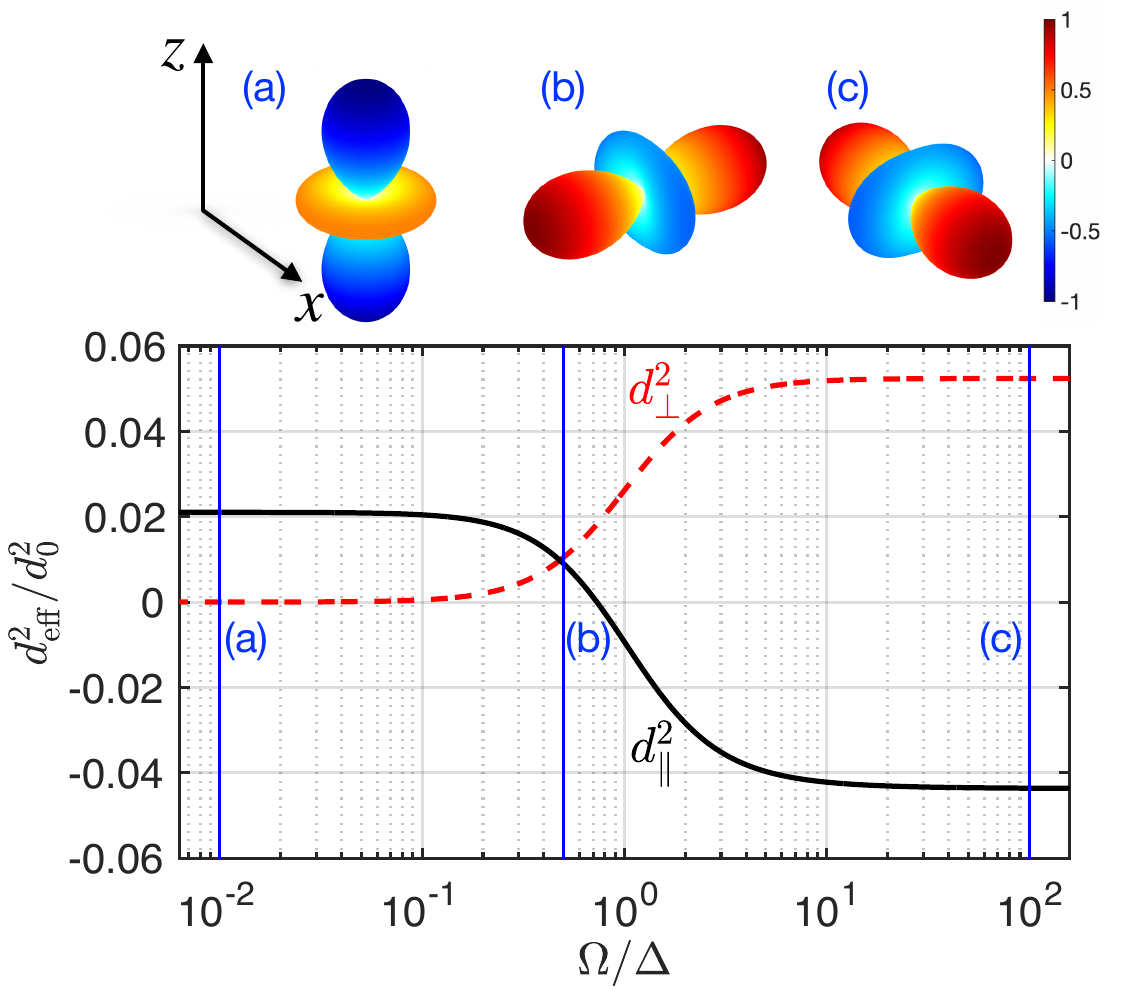}
    \caption{ Effective squared dipole moments parallel (solid black) and orthogonal (dashed red) to the dc field direction, normalized by the squared molecular frame dipole moment $d_0^2$, as a function of the ratio between the microwave Rabi frequency and detuning.  
    The subplots above show the normalized angular dependence of the first-order dipole-dipole interactions at (a) $\Omega/\Delta = 0.01$, (b) $\Omega/\Delta = 0.5$, and (c) $\Omega/\Delta = 100$ respectively, when $\xi = 45^{\circ}$. }
    \label{fig:deff2_vs_delta}
\end{figure}

For a fixed dc field, the effective dipole moments above only depend on the ratio $\Omega/\Delta$, and are plotted in Fig.~\ref{fig:deff2_vs_delta}.
In the figure, we highlight three specific ratios (vertical blue lines) that approximately produce the conventional first-order dipole-dipole angular dependence (up to an overall sign), but oriented along the $\hat{\boldsymbol{z}}$, $\hat{\boldsymbol{y}}$, and $\hat{\boldsymbol{x}}$ directions labeled (a), (b) and (c) respectively. This re-orientation of interaction occurs without any change in the applied field orientation or polarization, but rather through tuning $\Omega$ or $\Delta$. 
For other ratios, the angular dependence need not be cylindrically symmetric along any one of the Cartesian axes and can be fully anisotropic. 

The form of the potential (\ref{eq:firstorder_interaction}) is similar to that with a single elliptically polarized microwave \cite{Deng23_PRL}, although the degree to which interactions can be manipulated while maintaining highly effective shielding is greater in our scheme. The symmetries of this interaction potential, and its consequences on quantum droplet formation in Bose-condensed gases have been explored in Ref.~\cite{Baillie26_PRR}. 
Varying the interaction anisotropy relative to the tweezer trap geometry provides a tuning knob on the interaction energy shifts required for molecule spilling, as we will see in Sec.~\ref{sec:results}.

In this study, we consider NaCs molecules and fix the static electric field at ${\rm E}_{\rm dc}^* = 2.366$ kV/cm ($d_0 {\rm E}_{\rm dc}^* = 3.254 B_{0}$) unless stated otherwise, where the F\"orster energy defect between the $\ket{ g; g }$ and $\ket{ F }$ states is $\Delta_F/(2\pi) = 12.387$ MHz. 
This choice of dc field is motivated by experimentally achievable Rabi frequencies of $\sim 10$ MHz, allowing dominantly F\"orster resonant shielding while allowing the ac dressing to appreciably modify intermolecular interactions~\cite{Wang26_arxiv}. 

The second-order interaction is derived to be
\begin{align} \label{eq:secondorder_interaction}
    V^{(2)}_{\rm dd}(\boldsymbol{r}; \xi)
    &=
    \frac{ C_{\rm dc}(\hat{\boldsymbol{r}}) }{ r^6 }
    +
    \frac{ \bar{C}_{\rm dc}(\hat{\boldsymbol{r}}) }{ r^6 }
    +
    \frac{ C_{\rm ac}(\hat{\boldsymbol{r}}; \xi) }{ r^6 }.
\end{align} 
with effective dispersion functions
\begin{subequations}
\begin{align}
    C_{\rm dc}
    &= 
    \left(
    \frac{ d_{g\rightarrow{l}} d_{g\rightarrow{u}} }{ 4 \pi \epsilon_0 }
    \right)^2
    \frac{ 2 \alpha_+^2 ( 1 - 3 \cos^2\theta )^2 }{ \hbar ( \Delta_F - \Delta + \sqrt{ \Delta^2 + \Omega^2 } ) },
    \\
    \bar{C}_{\rm dc}
    &=
    \left(
    \frac{ d_{g\rightarrow{l}} d_{g\rightarrow{e}} }{ 4\pi\epsilon_0 }
    \right)^2
    \frac{ 9 \alpha_{+}^2 \sin^2(2\theta) }{ 4 \hbar( \Delta_{\bar{F}}-\Delta + \sqrt{ \Delta^2 + \Omega^2 } ) },
    \\
    C_{\rm ac}
    &= 
    \left(
    \frac{ d_{g \rightarrow e} d_{e \rightarrow g} }{ 4 \pi \epsilon_0 }
    \right)^2
    \frac{ 9 c_+^2 c_-^4 \sin^4\theta }{ \hbar ( \Delta + \sqrt{ \Delta^2 + \Omega^2 } ) } \\
    &\quad\quad\quad\quad
    \quad\quad\quad\quad \times 
    \left[ 
    1 - \cos^2(2\phi) \sin^2(2\xi) 
    \right],  \nonumber 
\end{align}
\end{subequations}
where $\hbar\Delta_F$ and $\hbar\Delta_{\bar{F}}$ are the F\"orster energy defects (i.e. the energy difference) between the pair states $|{F}\rangle$ and $|{\bar{F}}\rangle$, from the state $| g; g \rangle$. The subscript labels $l$ and $u$ in the transition dipole moments correspond to the dressed $|\tilde{N}, m\rangle = |\tilde{0}, 0\rangle$ and $|\tilde{2}, 0\rangle$ states respectively.

The first two terms in Eq.~(\ref{eq:secondorder_interaction}) correspond to perturbations from the energetically closest dc-induced pair states, while the third is from the closest ac-induced pair state. 
Despite its non-trivial anisotropy, $V_{\rm dd}^{(2)}(\boldsymbol{r})$ is completely repulsive along all collisional orientations, as is required for shielding.
The total effective potential is then the sum of first- and second-order terms $V_{\rm eff}(\boldsymbol{r}) = V_{\rm dd}^{(1)}(\boldsymbol{r}) + V_{\rm dd}^{(2)}(\boldsymbol{r})$, which describes the elastic interactions accurately both with and without an ac field.  
We expect the effective potential to be valid when $\Delta_F, \Delta \gg \hbar^2 / (\mu a_d^2)$ with the effective dipole length $a_d = \mu \max\{ d_{\parallel}^2, d_{\perp}^2 \} / (4\pi\epsilon_0\hbar^2)$. 
A detailed derivation of $V_{\rm eff}(\boldsymbol{r})$ is provided in App.~\ref{app:effective_potential_derivation}.

\subsection{ All-ellipticity bound state free shielding }

\begin{figure}[ht]
    \centering
    \includegraphics[width=\linewidth]{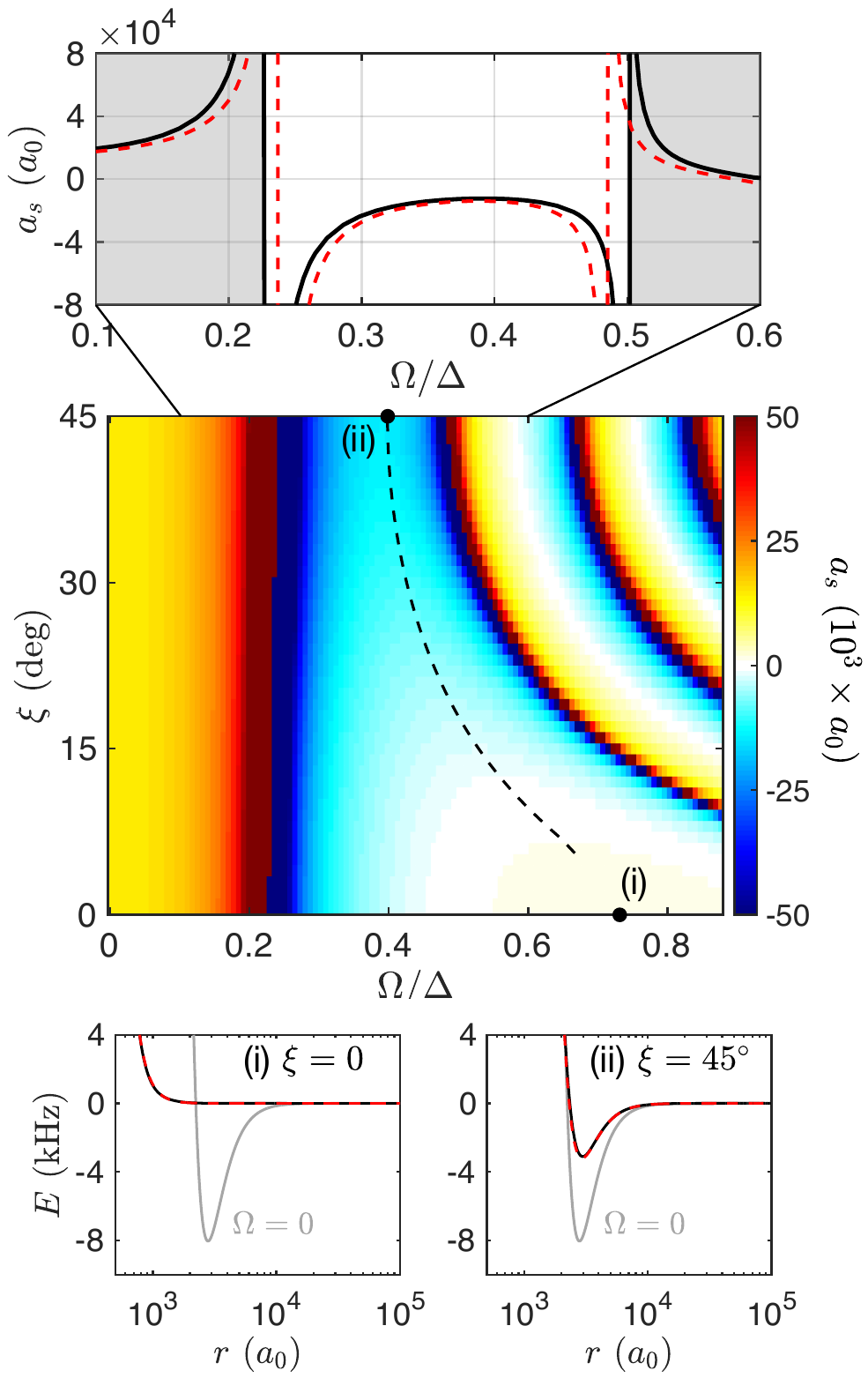}
    \caption{ Central panel: Real part of the  scattering length as a function of both Rabi frequency $\Omega$ and microwave ellipticity $\xi$. The color bar saturates at $\pm 50,000 a_0$ for clarity of presentation. 
    The dashed black curve is the ratio of optimal Rabi frequency $\Omega^{\star}$ to detuning $\Delta$ that minimizes the long-range potential well depth as a function of ellipticity angle.
    Upper panel: Real part of the scattering length, obtained from the multichannel calculations (solid black curve) and the effective potential (dashed red curve), as a function of Rabi frequency with $\xi = 45^{\circ}$. The ac field detuning is fixed at $\Delta = \Delta_F$, while the dc field is set at ${\rm E}_{\rm dc} = 2.366$ kV/cm. 
    Lower panels: (i) shielded adiabatic potential at the optimal ratio $\Omega^{\star}/\Delta$ with $\xi = 0$; (ii) shielded adiabatic potential at the optimal ratio $\Omega^{\star}/\Delta$ with $\xi = 45^{\circ}$. Both subplots compare the ac-dc(X) shielded potentials from the full multichannel calculation (solid black) and the effective potential (dashed red) to the shielded potentials with only dc shielding (i.e. $\Omega = 0$). }
    \label{fig:scatteringLength_vs_Omega}
\end{figure}

At high densities, a known loss mechanism in single-field shielded ultracold molecules is three-body recombination due to the presence of FL bound states.     
With controlled spilling relying on the stability of tightly confined molecular ensembles, a survey of field parameters that eliminate these FL states is necessary.
Such a survey is aided by Levinson's theorem, which connects the onset of bound states in the shielded potential with resonances in the $s$-wave scattering length $a_s$.  
As such, we employ close-coupling scattering calculations to obtain $a_s$, performed with an adaptive step size version of the Johnson log derivative propagation method \cite{Johnson73_JCP} at $10$ pK collision energy, including rotational states up to $N=4$, photon states $n = 0, -1, -2$, and collisional partial waves up to $L = 6$, with a universal loss short-range boundary condition \cite{Wang15_NJP} to account for short-range chemistry \cite{Hu2019_Science} or sticky dynamics \cite{Mayle12_PRA, Bause23_JPCA}.

In the case of ac-dc(X) shielding ($\xi = 45^{\circ}$), we find that there indeed exists a range of $\Omega/\Delta$ where no FL states are supported. 
This range is indicated by the unshaded region in the upper panel of Fig.~\ref{fig:scatteringLength_vs_Omega}, where on either side, a FL resonance occurs around $\Omega = 0.23 \Delta$ and $\Omega = 0.5 \Delta$.   
We compare the scattering length obtained from the effective potential (dashed red curve) to full close-coupling calculations (solid black curve) in Fig.~\ref{fig:scatteringLength_vs_Omega}. The good agreement in both the magnitude of $a_s$ and positions of the FL resonances showcases the accuracy of the effective potential in describing elastic interactions.

More generally, smaller ellipticity angles can increase the range of $\Omega$ that is bound state free, seen by the enlarging resonance-free (primarily blue and negative) region of $a_s$ in the central panel of Fig.~\ref{fig:scatteringLength_vs_Omega} as $\xi$ decreases \footnote{
Within the bound state free region, the effective dipole moments range around $0.06 d_0 < | d_{\perp} | < 0.1 d_0$ and $0.1 d_0 < | d_{\parallel} | < 0.13 d_0$, corresponding to $0.29 \:{\rm D} < | d_{\perp} | < 0.48 \:{\rm D}$ and $0.48 \:{\rm D} < | d_{\parallel} | < 0.63 \:{\rm D}$ in NaCs.
While these field-induced dipoles are relatively small, spilling to a single molecule per tweezer with our protocol would subsequently remove the requirement for ac-dc(X) shielding. The molecules can then be transferred back to their rotational ground state at a shifted dc field to induce a much larger dipole moment.
}.  
We find that for microwave ellipticities $\xi \gtrsim 10^{\circ}$, the scattering length remains negative even in the absence of a FL state, a result of the remnant long-range attractive potential well.
This attractive well that persists in the presence of ac-dc(X) fields is seen in subplot (ii), while complete removal of the well with circularly polarized microwaves is shown in subplot (i). In these subplots, the lowest shielded adiabatic curves obtained from multichannel diagonalization (solid black curves) are compared to those from the effective potential (dashed red curves), showing excellent agreement. 

Utilizing the effective potential, we also obtain the optimal ratio $\Omega^{\star}/\Delta$ that minimizes the depth of the long-range potential energy well for various values of $\xi$, plotted as the dashed black curve in the central panel of Fig.~\ref{fig:scatteringLength_vs_Omega}.
The well depth is ill-defined when the shielded potential is strictly above its asymptotic threshold for $\xi$ smaller than around $5^{\circ}$.
At $\xi = 45^{\circ}$, the shallowest attractive well requires a ratio of $\Omega^{\star}/\Delta \approx 0.4$ [subplot (ii)], while at $\xi = 0$, the optimal ratio $\Omega^{\star}/\Delta \approx 0.73$ is where the first-order dipole-dipole interaction vanishes [subplot (i)].

In the sections to follow, we will utilize the practically convenient limit of linear microwave polarization ($\xi = 45^\circ$), having just shown a broad operating range of $\Omega/\Delta$ over which three-body recombination due to the presence of FL bound states is suppressed.

\subsection{ Calculated rate coefficients } 

Non-adiabatic transitions and tunneling past the shielding barrier can still occur, leading to exothermic loss from the tweezer and limiting the efficacy of shielding.   
With the scheme reliant on multiple molecules co-trapped in a single tweezer, the stability of the system is paramount. The loss in such a many-molecule system can be inferred from scattering calculations that provide both elastic and inelastic collision rate coefficients $\beta_{\rm el} = \sigma_{\rm el} v_c$ and $\beta_{\rm loss} = \sigma_{\rm loss} v_c$ respectively, where $\sigma_{\rm el}$ and $\sigma_{\rm loss}$ are the corresponding integral cross sections, and $v_c = \sqrt{ 2 E_c / \mu }$ is the collision velocity.   

\begin{figure}[ht]
    \centering
    \includegraphics[width=\linewidth]{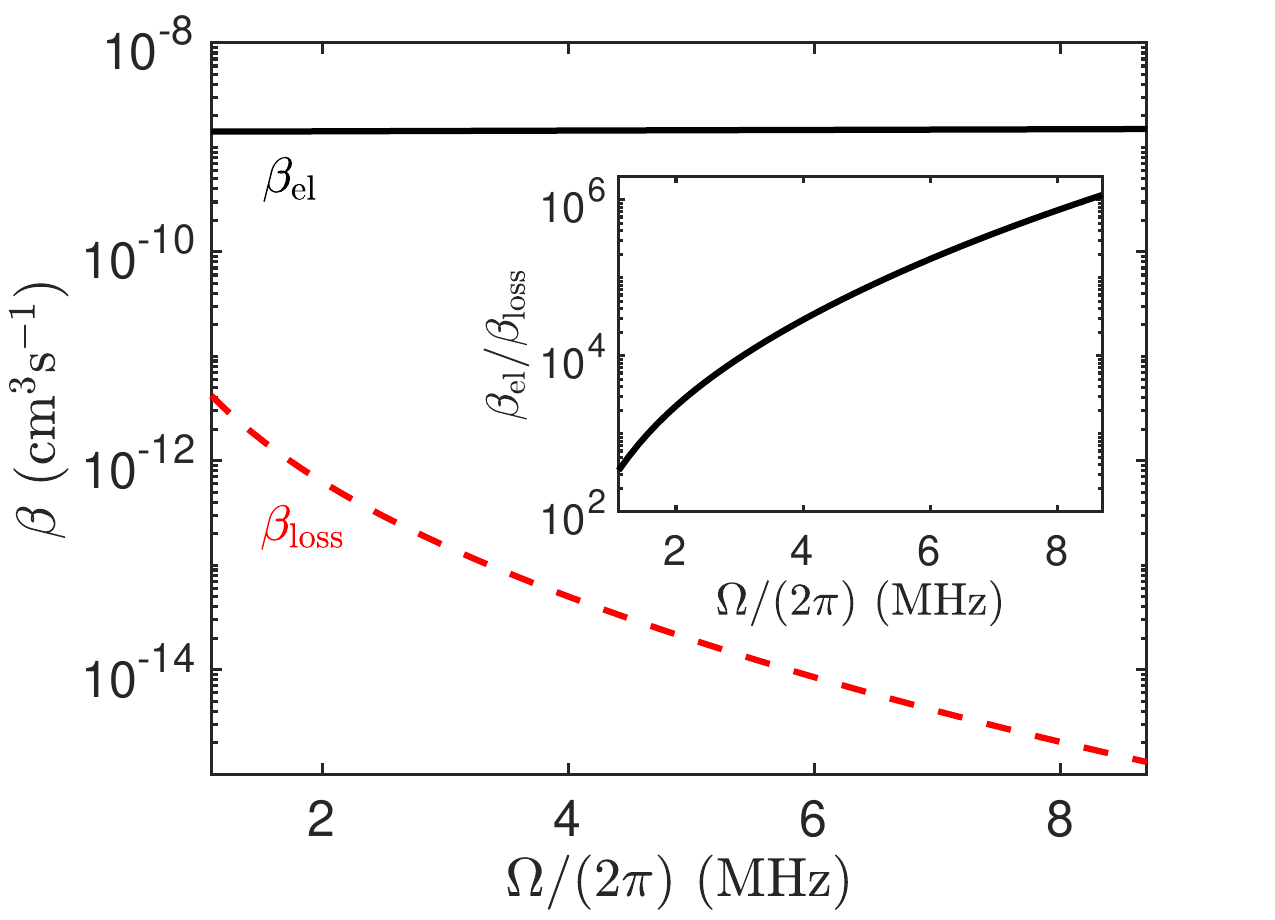}
    \caption{ Elastic (solid black curve) and total loss (dashed red curve) rate coefficients as a function of Rabi frequency at collision energy $E_c/k_B = 100$ nK. 
    The ac field is detuned such that $\Omega = 0.4 \Delta$, while the dc field is set at ${\rm E}_{\rm dc} = 2.366$ kV/cm. }
    \label{fig:rateCoefficient_vs_Omega}
\end{figure}

The  rate coefficients over a range of $\Omega$ are plotted in Fig.~\ref{fig:rateCoefficient_vs_Omega}, at a collision energy of $E_c/k_B = 100$ nK ($E_c/h = 2.084$ kHz), and at $\Omega = 0.4\Delta$ with $\Delta = \Delta_F$, ensuring no bound states.
It is favorable to have $\Delta \geq \Delta_F$ to minimize microwave-induced inelastic collisional losses. 
Elastic-to-loss rate ratios of $> 10^6$ can be achieved even with modest Rabi frequencies of $\Omega/(2\pi) < 10$ MHz, promising long molecular lifetimes for our tweezer loading protocol and theoretically predicted loss rate coefficients below what is currently achieved in bulk gas two-field shielding experiments ($\approx 10^{-14}$ cm$^3$/s) \cite{Shi26_NatPhys, Yuan25_arxiv, Biswas2026_arxiv}.

While strong tweezer confinement might mean that elastic rate coefficients calculated assuming free-space completed collisions
are inaccurate, inelastic collisions will increase the kinetic energy of a molecular pair by several MHz. With tweezer trap depths of $< 100$ kHz, the calculated loss rate coefficients should serve as good estimates since the molecules essentially propagate as free particles after an inelastic collision, and are primarily coupled to the elastic channel within a region smaller than the tweezer width.
We also compute the loss rate as a function of collision energy in App.~\ref{app:higher_energy_lossrates}, serving as an estimate of the two-body loss rate in tweezers with more than two molecules.  

As an added point of comparison, trapped loss rates can be computed using Fermi's golden rule \cite{Napolitano94_PRL, Julienne97_JRNIST} with the trapped molecular wavefunctions from imaginary time evolution (see Sec.~\ref{subsec:spilling}). We perform this analysis in App.~\ref{app:Fermis_golden_lossrates}. 
The Fermi's golden rule method makes a two-channel simplification and gives comparable or lower loss rates at the field parameters considered, so we retain use of the loss rates obtained from scattering calculations to provide conservative estimates.

\section{ Unit-loading of tweezers \label{sec:results} }

\subsection{ Interaction-induced spilling \label{subsec:spilling} }

The interactions engineered between two molecules from ac-dc(X) shielding act to stabilize the molecules against inelastic loss, while providing an in-tweezer interaction shift $\Delta E$, both of which are essential to the spilling process.
This energy shift increases the tunneling rate for the molecules in that site, and once one molecule tunnels out, the interaction shift is no longer present and tunneling is suppressed. This interaction-induced separation of tunneling timescales was experimentally demonstrated with distinguishable fermionic atoms \cite{Zurn12_PRL}, but not yet for the bosonic species that motivates the following analysis.

\begin{figure}[t]
    \centering 
    \includegraphics[width=\linewidth]{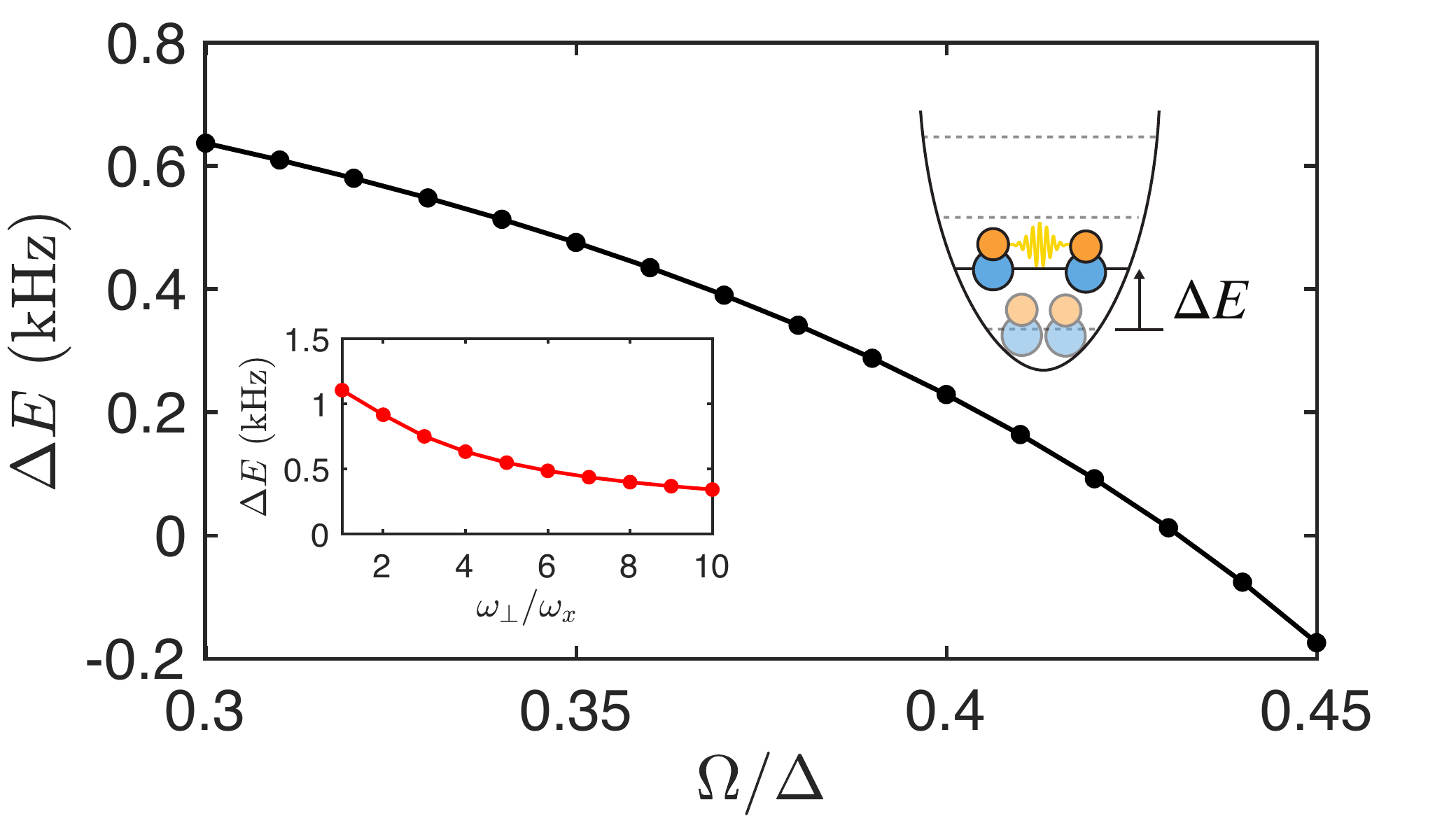} 
    \caption{ Interaction shift as a function of Rabi frequency at fixed detuning $\Delta = 1.3\Delta_F$, in a cylindrically symmetric trap with $\omega_x/(2\pi) = 0.5$ kHz and $\omega_{\perp}/(2\pi) = 5$ kHz. 
    The bottom left inset shows the interaction shift as a function of the trap aspect ratio with fixed $\bar{\omega}/(2\pi) = 3$ kHz. An illustration of the two-molecule interaction shift is given in the top right inset. }
    \label{fig:Eshift_vs_omega}
\end{figure}

To model the final two-to-one molecule isolation event,  we consider two molecules near the tweezer potential minimum, well described by a harmonic potential in their center-of-mass frame:
\begin{align}
    V_{\rm trap}(\boldsymbol{r})
    &=
    \frac{ 1 } { 2 }
    \mu
    \left[
    \omega_x^2 x^2
    +
    \omega_{\perp}^2 ( y^2 + z^2 )
    \right], 
\end{align}
where $\omega_{\perp}$ and $\omega_{x}$ are the harmonic frequencies.  
The tweezer light is taken to propagate along $x$, so that the trap is cylindrically symmetric about $\hat{\boldsymbol{x}}$. 
{By utilizing trap frequencies that already account for the spilling field gradient, the first-order correction due to trap anharmonicities is found to change $\Delta{E}$ by $< 5\%$ (see App.~\ref{app:anharmonic_corrections} for details). We account for these corrections with conservative estimates for $\Delta{E}$ in the analysis of spilling below. } 

The interaction-induced shift in energy $\Delta{E}$ between ac-dc(X) dressed molecules, both occupying the tweezer ground state, is then taken relative to the bare harmonic oscillator ground state energy:
\begin{align}
    \Delta{E}
    &=  
    \langle \psi_0 | 
    H 
    | \psi_0 \rangle   
    - 
    \frac{\hbar}{2} 
    \sum_i \omega_i
\end{align}
where $\ket{ \psi_0 }$ is the ground state of the Hamiltonian $H = {\cal H}_A + {\cal H}_B + \boldsymbol{p}^2/(2\mu) + V_{\rm trap}(\boldsymbol{r}) + V_{\rm eff}(\boldsymbol{r})$ in the center-of-mass frame of molecules $A$ and $B$. 
Due to the anisotropies in both the intermolecular interactions and the trap, we utilize numerical imaginary time evolution \cite{Lehtovaara07_JCP} on a three-dimensional grid to compute $\ket{\psi_0}$. 
We vary the resolution and size of the grid until we achieve convergence of the ground state energy at the sub-hertz level.
Fixing the trap frequencies as $\omega_x/(2\pi) = 0.5$ kHz and $\omega_{\perp}/(2\pi) = 5$ kHz, we plot the interaction shift as a function of $\Omega$ for a constant $\Delta = 1.3\Delta_F$ in Fig.~\ref{fig:Eshift_vs_omega}.
There are no FL bound states in the range of $\Omega$ plotted. 
The bottom left inset also shows that for a fixed geometric mean trap frequency $\overline{\omega}/(2\pi) = 3$ kHz, $\Delta{E}$ further increases in more isotropic tweezer potentials, achievable with a higher NA objective.

The interaction shift in Fig.~\ref{fig:Eshift_vs_omega} must be sufficiently large to separate the timescales of the first and second molecule tunneling upon application of the electric field gradient. 
With three-body loss strongly suppressed by bound-state-free shielding, we introduce a minimal probabilistic model for the spilling process to quantify the single-molecule isolation infidelity with uncorrelated boson tunneling \cite{Hunn13_PRA}. This model accounts for three competing processes starting from a doubly-occupied tweezer: (A) tunneling of one molecule from the doubly-occupied tweezer, (B) tunneling of the single remaining molecule, and (C) two-body loss. Each process is modeled as an exponential decay with a given timescale \footnote{Worth noting is that a two-body loss event of only two molecules in each tweezer is uncorrelated with the presence of other filled tweezers. Therefore, the number of molecules $N$ in an array of doubly filled tweezers follows the rate law $\dot{N} = -\tau_C^{-1} N$ with a constant $\tau_C$, the solution of which is $N(t) = N(0) e^{-{t/\tau_C}}$ rather than the solution of a two-body rate equation. }. 
Assigning the corresponding timescales $\tau_A, \tau_B,$ and $\tau_C$, we compute the probability $P_1(t)$ (detailed in App.~\ref{app:prob_model_description}) of obtaining a single molecule from a doubly-occupied tweezer after allowing tunneling to occur for a time $t$ by integrating over all of the times at which the first molecule tunnels out without (B) and (C) occurring:
\begin{align}
    P_1(t) = \int_{0}^{t} \frac{1}{\tau_A}
    e^{-\frac{t'}{\tau_A}} 
    e^{-\frac{t-t'}{\tau_B}} 
    e^{-\frac{t'}{\tau_C}}dt'.
\end{align}
To quantify these timescales, we select an optical tweezer with initial trapping frequencies of $\omega_{\perp}/(2\pi) = 5$ kHz and $\omega_{x}/(2\pi) = 1$ kHz. Application of a suitable spilling gradient reduces the axial trap frequency from $1$ kHz to approximately $0.5$ kHz without significantly changing the radial trap frequency (App.~\ref{app:spilling_potential}). 
Thus, the relevant ac-dc(X) interaction shift for this parameter set can be taken from Fig.~\ref{fig:Eshift_vs_omega}. 
Applying this gradient adiabatically then ensures that the molecules remain in the tweezer ground state with high probability, before significant tunneling dynamics occurs. 
The tunneling timescales $\{\tau_A, \tau_B\}$ are estimated at the final static spilling gradient using the interaction shift and the Hill-Wheeler approximation \cite{Kemble_PhysRev.48.549_1935, HillWheeler_PhysRev.89.1102_1953} for transmission through the near-parabolic barrier formed by applying a linear gradient to the optical tweezer potential (App.~\ref{app:tunneling_model}). 
The two-body loss timescale $\tau_C \approx 6.063$ s is estimated with the relevant two-body loss rate coefficient and a conservative estimate of the two-molecule density (App.~\ref{app:twobodyloss}).

\begin{figure}[ht]
    \centering
    \includegraphics[width=\linewidth]{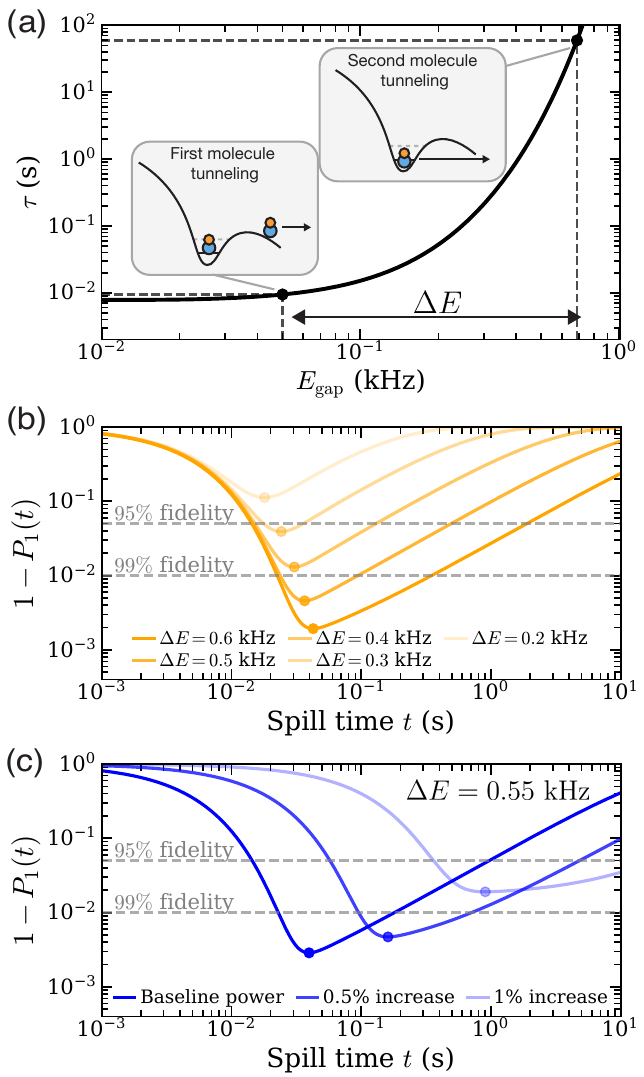}
    \caption{Tunneling and single molecule isolation. (a) Tunneling time constants are derived as a function of the energy gap from the top of a tilted tweezer for the case of NaCs molecules. The interaction shift, $\Delta E$, from ac-dc(X) shielding creates a separation of tunneling timescales for tunneling of the first and second molecule. (b) The infidelity of single molecule isolation for a range of interaction shifts for the tweezer parameters stated in the main text. Single molecule isolation fidelity of $> 99\%$ is achieved for interaction shifts $\Delta E > 0.5$ kHz. 
    (c) Single-molecule isolation fidelity at the fixed interaction shift $\Delta E = 0.55$ kHz for the baseline tweezer power used in panels (a) and (b) and for powers increased by 0.5\% and 1\% relative to the baseline. }
    \label{fig:tunneling_stats}
\end{figure}

Fig.~\ref{fig:tunneling_stats} reveals the results of this model: the interaction shift separates the tunneling timescales of the first of the two molecules ($\tau_A$) and the single remaining molecule ($\tau_B$) by nearly four orders of magnitude and the two-body loss rate is sufficiently slow to allow for $>99$\% single molecule isolation fidelity at the optimal spilling time for interaction shifts $\Delta E > 0.5$ kHz. 
In Fig.~\ref{fig:tunneling_stats}(a), $E_{\rm{gap}}$ is defined as the energy difference between the trapped molecular pair energy and the barrier maximum. 
The range of interaction shifts displayed in Fig.~\ref{fig:tunneling_stats}(b) is accessible by tuning the microwave Rabi frequency as shown in Fig.~\ref{fig:Eshift_vs_omega}. A larger interaction shift leads to a larger difference in tunneling timescales between the first and second molecules, resulting in a higher spilling fidelity. Fig.~\ref{fig:tunneling_stats}(c) contains the fidelity results for the interaction shift $\Delta E = 0.55$ kHz. {This choice is a representative conservative interaction shift that lies below the maximum computed shift of  $\Delta E = 0.6372$ kHz} shown in Fig.~\ref{fig:Eshift_vs_omega}, for the case of $\Omega/\Delta = 0.3$ in a tweezer with a baseline power leading to trap frequencies of $\omega_{x}/(2\pi) = 0.5$ kHz and $\omega_\perp/(2\pi) = 5$ kHz. Additional curves are included to indicate the sensitivity of the spilling protocol to tweezer power, discussed in the following section.

For tweezers initially occupied by $N \geq 3$ molecules, the energy shift associated with the removal of a molecule is $\approx (N-1)\Delta{E}$. 
The resulting tunneling timescale for higher occupancies is faster than for double occupancy at a given barrier height. Sites starting with $N>2$ molecules cascade down to $N = 2$ within the early stages of spilling, so that the major consideration is indeed the two-to-one isolation fidelity.

\subsection{ Extension to enhanced loading in arrays }

For simultaneous unit-loading of molecules into an array of hundreds of tweezers, the conditions for ac-dc(X) shielding and efficient spilling must all be satisfied simultaneously across the array. The critical experimental parameters are the uniformities of the dc field, the ac field, and the optical tweezer beam intensities.  
We detail major practical considerations for implementing this loading scheme across a full optical tweezer array.

Firstly, the ideally uniform dc field generated by electrodes in real systems is prone to small variation in ${\rm E}_{\rm dc}$ across different tweezer sites in the array.  
We show in App.~\ref{app:boundstate_dc_sensitivity} that there is a large window around ${\rm E}_{\rm dc}^*$ for which ${\rm E}_{\rm dc}$ can change while maintaining both F\"orster resonant shielding and no FL bound states for two- and three-body loss suppression. Specific geometrical arrangements of four or eight electrodes can be used to reduce the curvature of an applied dc field, resulting in high E-field uniformity and thus enabling increased array sizes for a given operating window around ${\rm E}_{\rm dc}^*$ for shielding \cite{Gempel2016-yp}. In App.~\ref{app:boundstate_dc_sensitivity}, the operating window for shielding spans $\pm 2.8$ V/cm for the case of NaCs molecules, which dictates a fractional field tolerance $\frac{\delta \rm E_{dc}}{\rm E_{dc}} \approx 10^{-3}$ given the shielding field of $2.366$ kV/cm. Current ultracold molecule experiments with high-voltage electrodes have demonstrated stability on the part-per-million level \cite{Tobias2022Thesis},  satisfying this constraint.

In the ac-dc(X) scheme, microwaves will address the $|\tilde{N} = 1, m = 0\rangle$ to $|\tilde{N} = 2, m = \pm 1\rangle$ rotational transition. For NaCs,
this transition frequency is $6.2186$ GHz at the shielding field $E = 2.366$ kV/cm. To quantify wavelength-scale microwave inhomogeneity, we consider the case in which the microwave interferes with a perfect counter-propagating co-polarized reflection to produce a standing wave along one of the array axes. Assuming the standing wave antinode is positioned at the array center and that the array spans 200 $\mu m$, the fractional change in Rabi frequency is computed to be $\frac{\delta \Omega}{\Omega} \approx 10^{-4}$ (App.~\ref{app:microwave_uniformity}). This fractional change is not expected to limit single-molecule isolation across the array given the broad operating window for the ratio of the Rabi frequency to the detuning $\Omega/\Delta$ presented in Fig.~\ref{fig:scatteringLength_vs_Omega} and the slow variation of the interaction shift as a function of Rabi frequency shown in Fig.~\ref{fig:Eshift_vs_omega}. Apparatus-dependent contributions to microwave non-uniformity can be characterized and minimized experimentally.

Finally, homogeneity of optical tweezer intensities is essential for maintaining high single molecule isolation fidelity.  
Intensity variation of optical tweezers in an array generated by a spatial light modulator has been controlled to a relative standard deviation of 0.3\% \cite{Chew_UltrapreciseTweezers_PhysRevA.110.053518}. Slightly relaxing this uniformity figure, Fig.~\ref{fig:tunneling_stats}(c) details the cases where there is a 0.5\% and 1\% power increase among sites in the tweezer array compared to the baseline tweezer power used for the calculations in Fig.~\ref{fig:tunneling_stats}(a,b). The electric field gradient is held constant; therefore the effect of this small power increase is that both the interaction-shifted and isolated single molecule states are energetically further from the barrier maximum. This larger energy gap leads to increased tunneling timescales and ultimately a different optimal spill time compared to the baseline case. Even with a static spilling gradient, a spill time can be chosen such that the per-site fidelity exceeds 95\% for the baseline and increased-power tweezers. 

The above analysis of tolerances suggests that this protocol is compatible with the field and trap inhomogeneities of realistic experimental systems, indicating that it can be implemented to produce high filling-fraction arrays of molecules.

\section{ Summary and outlook \label{sec:conclusions} }

We have outlined a protocol to efficiently isolate single molecules within optical tweezers in the presence of static electric and microwave fields. 
Starting with multiple molecules in each tweezer, the carefully applied ac-dc fields provide seconds-long stability against inelastic molecular loss, and allow for a gradual spilling of molecules from the tweezers in an electric field gradient until only one is remains. 
A practical appeal of this field dressing protocol is that it is robust to microwave polarization, a flexibility not afforded in previous microwave shielding methods.

A statistical analysis that accounts for molecular loss and experimentally realizable tolerances demonstrates that unit-loading of molecules in a tweezer array of initially doubly occupied sites could exceed $99\%$ fidelity in a single isolated site, and $95\%$ fidelity per tweezer across an array.
This prediction constitutes a leap toward realizing scalable optical tweezer arrays of polar molecules for quantum science applications.  
This fidelity can be further improved in practice with laser and dc field stability improvements. The spilling model used here assumes a constant gradient for the entire spilling procedure. To account for array inhomogeneity, an improvement may come from optimizing a time-dependent sequence for the applied field gradient used for spilling.

More broadly, there has been growing interest in the design of deterministic loading protocols for molecular tweezer arrays~\cite{Walraven24_PRL, Karman26_arxiv, Baldock26_arxiv}. 
The present work establishes a protocol for unit loading that can be readily adapted to dipolar bosonic molecules beyond the $^{1}\Sigma$ NaCs system studied here.
Indeed, the ac-dc(X) shielding technique is applicable to a broad range of dipolar diatomic species, including those with $^{2}\Sigma$ electronic ground states \cite{Quemener16_PRA, Wang26_arxiv, Ho26_PRR}.     
By leveraging collisional shielding and controlled spilling dynamics in optical tweezers, this protocol paves a path towards highly filled molecular arrays for quantum science.

\begin{acknowledgments}

RRWW and HRS acknowledge support by ITAMP with funding from the National Science Foundation. 
CHN, CW, AY, LD, and KKN acknowledge funding from the Gordon and Betty Moore Foundation (DOI 10.37807/GBMF11558), and the Brown Investigator Award, a program of the Brown Institute for Basic Sciences at the California Institute of Technology (S724889).

\end{acknowledgments}

\appendix

\section{ Derivation of effective potential \label{app:effective_potential_derivation} }

This section derives the effective potential for ac-dc(X) shielded molecules. 
In the truncated basis of Tab.~\ref{tab:basis_set}, the non-interacting Hamiltonian is
\begin{align}
    & \boldsymbol{{\cal H}} / \hbar = \\
    & \begin{pmatrix}
        0 & 0 & 0 & -\frac{\Omega}{2} & \frac{\Omega}{2} & 0 & 0 & 0 \\ 
        0 & -\Delta_F & 0 & 0 & 0 & 0 & 0 & 0 \\
        0 & 0 & -\Delta_{\bar{F}} & 0 & 0 & 0 & 0 & 0 \\
        -\frac{\Omega}{2} & 0 & 0 & -\Delta & 0 & \frac{\Omega}{2\sqrt{2}} & 0 & -\frac{\Omega}{2} \\
        \frac{\Omega}{2} & 0 & 0 & 0 & -\Delta & -\frac{\Omega}{2\sqrt{2}} & \frac{\Omega}{2} & 0 \\
        0 & 0 & 0 & \frac{\Omega}{2\sqrt{2}} & -\frac{\Omega}{2\sqrt{2}} & -2 \Delta & 0 & 0 \\
        0 & 0 & 0 & 0 & \frac{\Omega}{2} & 0 & -2 \Delta & 0 \\
        0 & 0 & 0 & -\frac{\Omega}{2} & 0 & 0 & 0 & -2 \Delta
    \end{pmatrix}, \nonumber
\end{align}
which can be diagonalized to give the ac-dc dressed eigenstates and eigenenergies:
\begin{align} 
    \begin{tabular}{l|c}
        dressed state & energy \\
        \hline
        $|{ +; + }\rangle_S$ & $2 \varepsilon_{+}$ \\
        $|{ F }\rangle\ket{ 0 }$ & $-\hbar\Delta_{F}$ \\
        $|{ \bar{F} }\rangle\ket{ 0 }$ & $-\hbar\Delta_{\bar{F}}$ \\
        $|{ +; \emptyset }\rangle_S$ & $\varepsilon_{+} - \hbar\Delta$ \\
        $|{ -; + }\rangle_S$ & $\varepsilon_{-} + \varepsilon_{+}$ \\
        $|{ -; \emptyset }\rangle_S$ & $\varepsilon_{-} - \hbar\Delta$ \\
        $|{ \emptyset; \emptyset }\rangle_S$ & $-2\hbar\Delta$ \\
        $|{ -; - }\rangle_S$ & $2 \varepsilon_{-}$
    \end{tabular}.
\end{align}
We note that both $|{ \tilde{2}, \pm 1 }\rangle$ states have the same energy, but we only include one of them, $|\bar{ F }\rangle = |{ \tilde{2}, +1 }\rangle$ in our truncated basis set. 
This simplification is valid due to the fact that $\Delta_{\bar{F}} \gg \Delta_{F}$ close to the F\"orster resonance.
Therefore, in the subspace of these two degenerate states with adiabatic Hamiltonian
\begin{align}
    \boldsymbol{{\cal H}}_{\rm deg}
    &=
    ( V_{\bar{F}} - \hbar\Delta_{\bar{F}} )
    \begin{pmatrix}
        1 & 0 \\
        0 & 1
    \end{pmatrix}
    +
    \begin{pmatrix}
        0 & \bar{W}(\boldsymbol{r}) \\
        \bar{W}^*(\boldsymbol{r}) & 0
    \end{pmatrix},
\end{align}
the off-diagonal dipolar interactions (defined below) only cause the symmetric state of the two hybridized states to be shifted closer to $\ket{g; g}$ as $r$ decreases.
The other state remains far detuned and less relevant to the elastic scattering.  
Therefore, we only keep $|{ \tilde{2}, + 1 }\rangle$ in the truncated basis with the same dominant field-seeking character as the symmetric state.

Dipolar interactions are mediated by the induced and transition dipole moments 
\begin{align}
    d_{a \rightarrow a'} 
    &=
    d_0 
    \sum_{ N', N }
    \langle \tilde{N}', m' | N', m' \rangle
    \langle N, m | \tilde{N}_, m \rangle \nonumber\\
    &\quad\quad\quad \times
    \sqrt{ (2 N'_B + 1) (2 {N_B} + 1) } \nonumber\\
    &\quad\quad\quad \times
    \begin{pmatrix}
        N' & 1 & {N} \\
        -m' & \Delta{m} & m
    \end{pmatrix}
    \begin{pmatrix}
        N' & 1 & {N} \\
        0 & 0 & 0
    \end{pmatrix},
\end{align}
where indices label states $a, a' \in \{ g, e, u, l \}$ as in Tab.~\ref{tab:basis_set}, and with $|l\rangle = |\tilde{0},0\rangle$ and $|u\rangle = |\tilde{2},0\rangle$.
We note that $d_{e} = d_{\bar{e}}$ and $d_{e \rightarrow g} d_{g \rightarrow e} = d_{\bar{e} \rightarrow g} d_{g \rightarrow \bar{e}}$.

Now computing the dipole-dipole interaction matrix elements in the symmetrized dc-dressed basis of Tab.~(\ref{tab:basis_set}), the diagonal matrix elements are 
\begin{subequations}
\begin{align}
    V_{gg}
    &=
    \frac{ d_g^2 }{ 4 \pi \epsilon_0 }
    \frac{ 1 - 3 \cos^2\theta }{ r^3 },
    \\
    V_{F}
    &=
    \frac{ d_{l} d_{u} + d_{l\rightarrow{u}} d_{u\rightarrow{l}} }{ 4 \pi \epsilon_0 }
    \frac{ 1 - 3 \cos^2\theta }{ r^3 },
    \\
    V_{\bar{F}}
    &=
    -\frac{ ( d_{l} d_{e} + d_{l\rightarrow{e}} d_{e\rightarrow{l}} / 2  )}{ 4 \pi \epsilon_0 }
    \frac{ 1 - 3 \cos^2\theta }{ r^3 }, 
    \\
    V_{eg}
    &=
    V_{\bar{e}g} =
    -\frac{ d_{g} d_{e} + d_{g\rightarrow{e}} d_{{e}\rightarrow{g}} / 2 }{ 4 \pi \epsilon_0 }
    \frac{ 1 - 3 \cos^2\theta }{ r^3 },
    \\
    V_{ee} 
    &= 
    V_{e\bar{e}} = V_{\bar{e}\bar{e}}
    =
    \frac{ d_{e}^2 }{ 4 \pi \epsilon_0 }
    \frac{ 1 - 3 \cos^2\theta }{ r^3 },
\end{align}
\end{subequations}
utilizing the notation $V_{ab} = \langle a; b | V_{\rm dd} | a; b \rangle$ with $a,b \in \{ g, e, F \}$.
Similarly, the non-trivial off-diagonal elements are
\begin{subequations}
\begin{align}
    U 
    &= 
    \langle \bar{e}; g |
    V_{\rm dd}
    | e; g \rangle
    =
    \frac{ d_{e \rightarrow g} d_{g \rightarrow \bar{e}} }{ 4 \pi \epsilon_0 }
    \frac{ 3 e^{2 i \phi} \sin^2\theta }{ 2 r^3 }, \\
    W 
    &= 
    \langle g; g |
    V_{\rm dd}
    | F \rangle
    =
    \langle F |
    V_{\rm dd}
    | g; g \rangle \nonumber\\
    &=
    -\frac{ \sqrt{ 2 } d_{g\rightarrow{u}} d_{g\rightarrow{l}} }{ 4 \pi \epsilon_0 }
    \frac{ 1 - 3 \cos^2\theta }{ r^3 }, \\
    \bar{W}
    &=
    \langle{g; g}| 
    V_{\rm dd}(\boldsymbol{r})
    |\bar{F}\rangle \nonumber\\
    &=
    -\frac{ (d_{g\rightarrow{e}}d_{g\rightarrow{l}} / 2) }{ 4\pi\epsilon_0 }
    \frac{ 3 e^{+ i \phi} \sin(2\theta) }{ r^3 }, \\
    X
    &=
    \langle{F}| 
    V_{\rm dd}(\boldsymbol{r})
    |\bar{F}\rangle \nonumber\\
    &=
    -\frac{ (d_{l} d_{e\rightarrow{u}} + d_{l\rightarrow{u}} d_{e\rightarrow{l}}) }{ 4 \pi \epsilon_0 \sqrt{2} }
    \frac{ 3 e^{ + i\phi } \sin(2\theta) }{ r^3 },
\end{align}
\end{subequations}
which allow us to write the dipole-dipole coupling matrix elements as
\begin{align}
    \boldsymbol{V}_{\rm dd} 
    &= 
    \begin{pmatrix} 
        V_{gg} & W & \bar{W} & 0 & 0 & 0 & 0 & 0 \\
        W & V_{F} & X & 0 & 0 & 0 & 0 & 0 \\
        \bar{W}^* & X^* & V_{\bar{F}} & 0 & 0 & 0 & 0 & 0 \\
        0 & 0 & 0 & V_{\bar{e}g} & U & 0 & 0 & 0 \\
        0 & 0 & 0 & U^* & V_{eg} & 0 & 0 & 0 \\
        0 & 0 & 0 & 0 & 0 & V_{e\bar{e}} & 0 & 0 \\
        0 & 0 & 0 & 0 & 0 & 0 & V_{ee} & 0 \\
        0 & 0 & 0 & 0 & 0 & 0 & 0 & V_{\bar{e}\bar{e}}
    \end{pmatrix}. 
\end{align}
The first- and second-order interaction potentials are then obtained through
\begin{subequations} \label{eq:Vdd1_Vdd2_X}
\begin{align}
    V_{\rm dd}^{(1)}(\boldsymbol{r})
    &=
    _S\bra{ +;+ }
    V_{\rm dd}(\boldsymbol{r}) 
    \ket{ +;+ }_S, \\
    V^{(2)}_{\rm dd}(\boldsymbol{r})
    &=
    \frac{ | \bra{0}\bra{ F } {V}_{\rm dd}(\boldsymbol{r}) \ket{ +; + }_S |^2 }{ 2\varepsilon_{+} - ( -\hbar\Delta_F ) } 
    \nonumber\\
    &\quad
    +
    \frac{ | _S\bra{ F_+ } {V}_{\rm dd}(\boldsymbol{r}) \ket{ +; + }_S |^2 }{ 2\varepsilon_{+} - ( -\hbar\Delta_{\bar{F}} ) }
    \nonumber\\
    &\quad  
    +
    \frac{ | _S\bra{ +; \emptyset } {V}_{\rm dd}(\boldsymbol{r}) \ket{ +; + }_S |^2 }{ 2\varepsilon_{+} - ( \varepsilon_{+} - \hbar\Delta ) },
\end{align}
\end{subequations}
where the second-order interaction only considers the three energetically closest states. The result is the potential in Eqs.~(\ref{eq:firstorder_interaction}) and (\ref{eq:secondorder_interaction}). 

It is straightforward to extend the derivation above to the case of elliptically polarized microwaves.

\section{ Estimating two-body loss with more molecules \label{app:higher_energy_lossrates} }

\begin{figure}[ht]
    \centering
    \includegraphics[width=\linewidth]{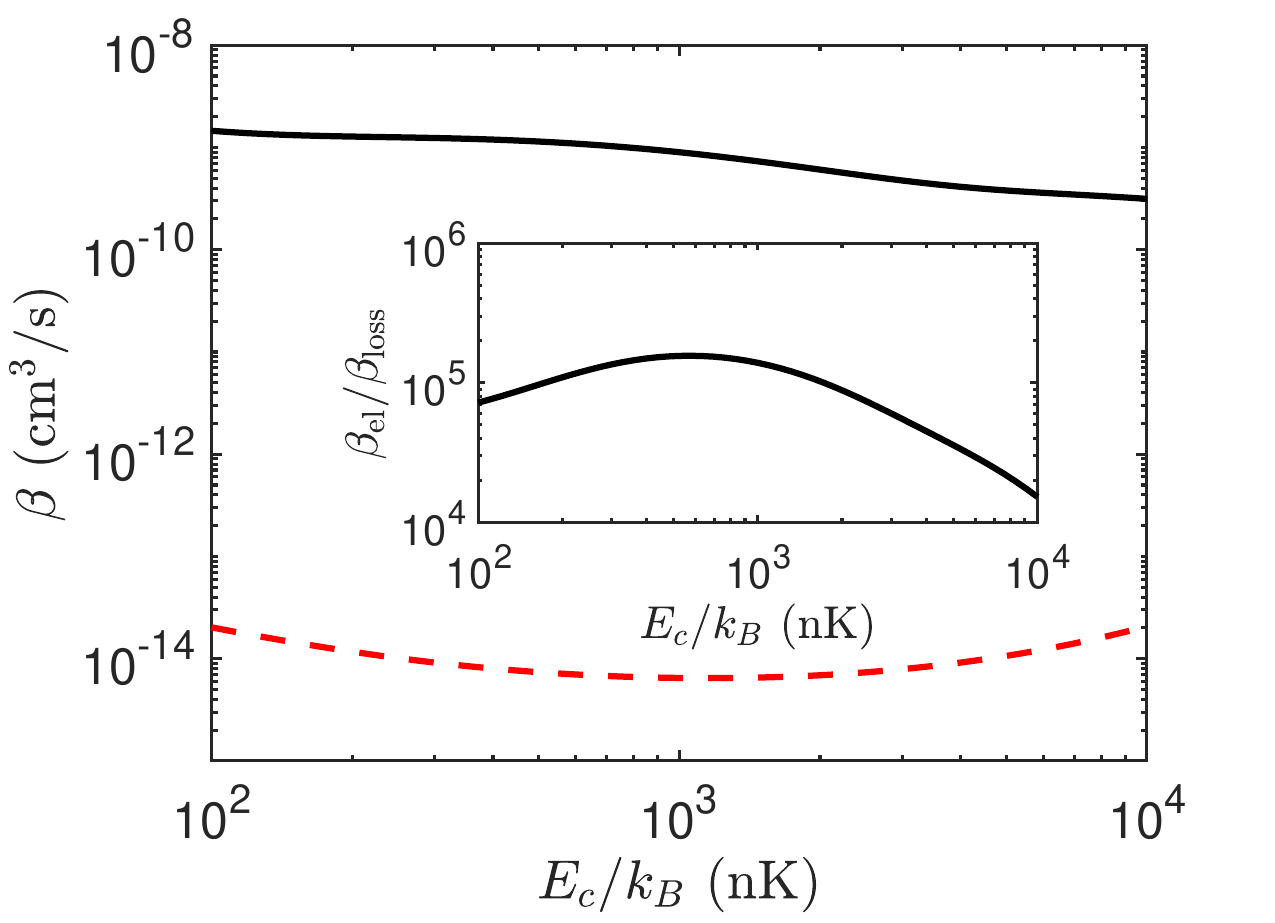}
    \caption{ Elastic (solid black) and loss (dashed red) rate coefficients as a function of collision energy, for $\Omega = 0.4\Delta = 2\pi \times 5$ MHz and $\Delta = \Delta_F$. The inset plots the ratio of elastic-to-loss rates as a function of collision energy. %
    }
    \label{fig:rateCoefficient_vs_collEnergy}
\end{figure}

When more than two molecules occupy a tweezer, strong dipolar interactions would likely render higher motional states of the tweezer energetically accessible, leading to a high probability of excited tweezer state occupation. 
Full quantum calculations of the loss rates in such a multi-band many-body system of field-dressed molecules are beyond the scope of this work, but we can provide estimates from two-body scattering calculations through the following argument.     
In a high density, low entropy sample of tweezer trapped molecules, ac-dc(X) shielding results in spatial regions where the molecular wavefunction is excluded due to the large intermolecular repulsive cores. 
This exclusion results in larger curvatures in the molecular wavefunctions, leading to an effective increase in kinetic energy of the sample.     
So as a proxy for determining the shielding efficacy at higher molecule numbers, we perform two-body scattering calculations at increasing collision energies as shown in Fig.~\ref{fig:rateCoefficient_vs_collEnergy}.
Assuming $\Omega = 0.4 \Delta = 2\pi \times 5$ MHz and $\Delta = \Delta_F$, we see that a high elastic-to-loss rate ratio of $\gtrsim 10^4$ is maintained even up to collision energies of 10 $\mu$K ($\approx$ 208 kHz).

\section{ Sensitivity of FL bound states to the dc field \label{app:boundstate_dc_sensitivity} }

\begin{figure}[ht]
    \centering
    \includegraphics[width=\linewidth]{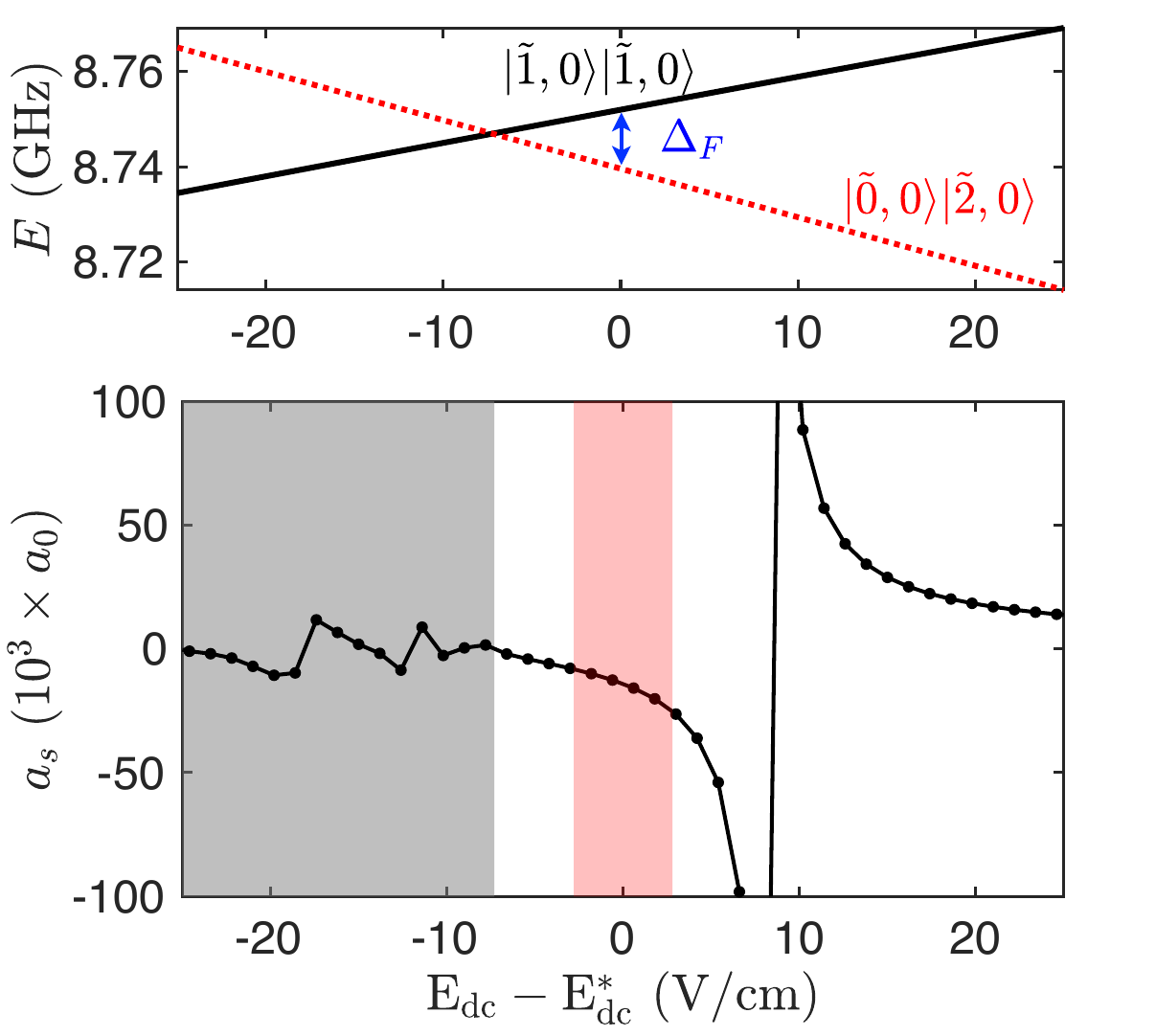}
    \caption{ Upper panel: Two-molecule dc dressed rotational states (labeled in plot) close to the F\"orster resonant crossing. The F\"orster energy defect $\Delta_F$ is labeled by the blue double-sided arrow.
    Lower panel: Real scattering length as a function of the dc field away from ${\rm E}_{\rm dc}^* = 2.366$ kV/cm. F\"orster resonant shielding is not effective when $|\tilde{0},0\rangle|\tilde{2},0\rangle$ is energetically above $|\tilde{1},0\rangle|\tilde{1},0\rangle$, indicated by the gray shaded region. }
    \label{fig:as_DCdensitivity}
\end{figure}

Fig.~\ref{fig:as_DCdensitivity} demonstrates that when varying the dc field within the window of operation across the tweezer array (well-bounded within red shaded region in the lower panel), the real scattering length shows no FL resonances while maintaining that $|\tilde{1},0\rangle|\tilde{1},0\rangle$ remains energetically above $|\tilde{0},0\rangle|\tilde{2},0\rangle$, necessary for F\"orster resonant shielding (see the upper panel).
A FL resonance appears to the right of this zone, while several other resonances appear to the left within the gray shaded region due to threshold crossings of field dressed rotational states. 
For the case of NaCs molecules, the operating window for shielding spans $\pm 2.8$ V/cm around the center field value.

\section{ Gradient-spilling in the tweezer \label{app:spilling_potential}}

For the spilling model presented in the main text, we assume that the trapping potential is formed from a Gaussian beam and that the molecules remain localized in the radial coordinates $y = z = 0$. In this case, the axial tweezer potential with a gradient arising from a dc Stark shift can be written in the following manner:
\begin{align}
     V(x) = -V_0\bigg(1 + \bigg(\frac{x}{x_R}\bigg)^2\bigg)^{-1} - d_g\frac{\partial \rm E_{dc}}{\partial x}x
\end{align}
where $x_R = \frac{\pi w_0^2}{\lambda}$ is the Rayleigh range. For this model, we assume that the wavelength is $\lambda = 1064$ nm and that the trap frequency aspect ratio is $\omega_{\perp}/\omega_x = 5$, which constrains the waist to $w_0 \approx 1.197 \mu$m. This is a reasonable waist to consider as the diffraction limit permits $w_0 \approx 0.61\lambda/\rm{NA}$ where $\rm{NA}$ is the numerical aperture of the microscope objective. At our chosen field of ${\rm E}_{\rm dc} = 2.366$ kV/cm, the dc field induced dipole moment is $d_g = -0.689$ D. 

The magnitude of the electric field gradient is chosen such that the energy gap between the barrier maximum and the lowest energy quasi-bound state is equal to the interaction shift plus an additional 50 Hz. This is verified by imposing infinite walls at a given distance on the $x$-axis and numerically solving the one-dimensional Schr\"{o}dinger equation. For the given trap parameters and dipole moment, this gradient is approximately 0.579 kV/cm$^2$ (with few-percent-level variation depending on the desired gap to the lowest quasi-bound state), which is an accessible field gradient in typical experimental platforms using high-voltage electrodes. 

The application of the gradient strongly decreases the effective trap frequency in the axial (spilling) direction. Due to the stronger confinement in the radial directions of the tweezer, a gradient of the same scale does not significantly reduce the trap frequency. In the case of a tweezer with original aspect ratio $\omega_{\perp}/\omega_x = 5$ with $\omega_{x}/(2\pi) = 1$ kHz and $\omega_{\perp}/(2\pi) = 5$ kHz, application of the aforementioned gradient reduces the axial trap frequency to $\omega_{x, \rm{spilled}}/(2\pi) = 0.505$ kHz and $\omega_{\perp, \rm{spilled}}/(2\pi) = 4.970$ kHz.

As a conservative estimate, interaction shifts are then computed in the spilled configuration of the tweezer assuming this new geometric mean trap frequency and aspect ratio. This consideration is crucial because estimating the interaction shift with smaller aspect ratios and higher geometric mean trap frequencies would dramatically overestimate the single molecule isolation fidelity. In Fig.~\ref{fig:Eshift_vs_omega}, we conservatively compute the interaction shift with $\omega_{x}/(2\pi) = 0.5$ kHz and $\omega_{\perp}/(2\pi) = 5$ kHz.

\section{ Anharmonic corrections to the interaction shift \label{app:anharmonic_corrections} }

The first-order correction to the interaction shift due to trap anharmonicities is derived here. 
To begin, we model the tweezer potential to be perfectly harmonic along $y$ and $z$ and to have a Lorentzian profile along $x$ where spilling occurs:
\begin{align}
    V(\boldsymbol{r}_1, \boldsymbol{r}_2) 
    &=
    \frac{ 1 }{ 2 }
    M \omega_{\perp}^2
    ( Y^2 + Z^2 )
    +
    \frac{ 1 }{ 2 }
    \mu \omega_{\perp}^2
    ( y^2 + z^2 ) \\ 
    &+
    2V_0
    -
    \frac{ V_0 }{ 1 + \left[ (X \pm x/2)/x_R \right]^2 }
    -
    2 d_g 
    \frac{\partial \rm E_{dc}}{\partial x}
    X. \nonumber
\end{align}
The center-of-mass and relative coordinates are defined as $\boldsymbol{R} = ( \boldsymbol{r}_1 + \boldsymbol{r}_2 ) / 2$ and $\boldsymbol{r} = \boldsymbol{r}_1 - \boldsymbol{r}_2$. 

To solve for the interaction shift in this anharmonic potential, we use the ground state obtained in a harmonic potential (with frequencies obtained from the Hessian around the local minima) and then treat the anharmonic terms in perturbation theory. To do so, we define the harmonic potential
\begin{align}
    V_{\rm harm}(\boldsymbol{r}, \boldsymbol{R})
    &=
    \frac{ 1 }{ 2 }
    M \omega_{\perp}^2
    ( Y^2 + Z^2 )
    +
    \frac{ 1 }{ 2 }
    M \omega_{x}^2
    X^2 \nonumber\\
    &\quad 
    +
    \frac{ 1 }{ 2 }
    \mu \omega_{\perp}^2
    ( y^2 + z^2 ) 
    +
    \frac{ 1 }{ 2 }
    \mu \omega_{x}^2
    x^2,
\end{align}
with which we define the anharmonic perturbative potential as
\begin{align}
    \delta{V}(\boldsymbol{r}, \boldsymbol{R})
    &=
    V(\boldsymbol{r}, \boldsymbol{R}) 
    -
    V_{\rm harm}(\boldsymbol{r}, \boldsymbol{R}) \nonumber\\
    &=
    2V_0
    -
    \frac{ V_0 }{ 1 + \left[ (X \pm x/2)/x_R \right]^2 }
    -
    2 d_g 
    \frac{\partial \rm E_{dc}}{\partial x}
    X \nonumber\\
    &\quad 
    -
    \frac{ 1 }{ 2 }
    M \omega_{x}^2
    X^2
    -
    \frac{ 1 }{ 2 }
    \mu \omega_{x}^2
    x^2.
\end{align}
Because the linear term in $X$ acts to shift the position of the potential minima by $\delta{X}$ 
and offset the potential minima by an amount $V_{\min}$, we simply include its effect by a shift of the coordinate frame $X \rightarrow X + \delta{X}$ and subtracting away the constant offset:
\begin{align}
    \delta{V}(\boldsymbol{r}, \boldsymbol{R})
    &\approx 
    2V_0
    -
    \frac{ V_0 }{ 1 + \left[ (X + \delta{X} \pm x/2)/x_R \right]^2 }
    -
    V_{\min}
    \nonumber\\
    &\quad 
    -
    \frac{ 1 }{ 2 }
    M \omega_{x}^2
    X^2
    -
    \frac{ 1 }{ 2 }
    \mu \omega_{x}^2
    x^2.
\end{align}
The approximation above is valid when $\delta{X}$ is much smaller than $x_R$, which is true in our case. 

With the perturbative terms only involving $x$ and $X$ coordinates, the expectation with the center-of-mass harmonic ground state wavefunction can be evaluated analytically:
\begin{align} \label{eq:COM_expectation}
    & \langle \Psi_0(\boldsymbol{R}) |
    \delta{V}(\boldsymbol{r}, \boldsymbol{R})
    | \Psi_0(\boldsymbol{R}) \rangle \\
    &=
    2 V_0 
    -
    \frac{V_0}{2}
    {\cal L}(2\delta{X} \pm  x)
    - 
    V_{\min}
    -
    \frac{ 1 }{ 4 } \hbar\omega_x
    -
    \frac{ 1 }{ 2 }
    \mu \omega_{x}^2
    x^2, \nonumber 
\end{align}
where 
\begin{align}
    {\cal L}(x)
    &=
    2 \sqrt{ \pi } 
    \left( \frac{x_R}{a_X} \right) 
    e^{ -\frac{ x^2 }{ 4 a_X^2 } } \nonumber\\
    &\quad \times
    e^{ \frac{ ( x_R \mp i x ) x_R }{ a_X^2 } }
    \left[ 1 \pm {\rm Erf}\left( \frac{ i x \mp 2 x_R }{ 2 a_X } \right) \right],
\end{align}
and $a_X = \sqrt{ \hbar/(M \omega_x) }$. 
Above, ${\rm Erf}(z)$ is the error function. 
The energy shift from the anharmonic perturbation is then computed by taking the expectation of Eq.~(\ref{eq:COM_expectation}) with the relative coordinate ground state $\psi_0(\boldsymbol{r})$ obtained from imaginary time evolution in Sec.~\ref{subsec:spilling}.

\section{Model to estimate tunneling timescales \label{app:tunneling_model}}
When the gradient is applied, tunneling will occur along the weakly-confined axial ($\hat{x}$) direction. Ref.~\cite{Hunn13_PRA} showed that sequential tunneling of each boson out of the trap into the continuum is highly favored over the simultaneous tunneling of both. By engineering the molecular interactions to ensure the interaction-shifted single-particle energy is significantly closer to the top of the barrier than that of an isolated single molecule, application of an electric field gradient can cause one molecule from the pair to spill out. To quantify the fidelity of this process, we impose the following model for extracting tunneling timescales:
\begin{align}
    \label{eq:tautimescale}
    \tau = (f_{\text{attempt}} \cdot T_{HW})^{-1}
\end{align}
where $f_{\text{attempt}}$ is the inverse of the semiclassical round-trip time to travel between the classical turning points in the tweezer and $T_{HW}$ is the transmission coefficient for tunneling under the Hill-Wheeler approximation \cite{Kemble_PhysRev.48.549_1935, HillWheeler_PhysRev.89.1102_1953}. The attempt frequency depends on the classical turning points $x_1, x_2$, the molecule mass $m$, and the energy in the tweezer $E$:
\begin{align}
     f_{\text{attempt}} =\bigg(\int_{x_1}^{x_2} \frac{2dx}{\sqrt{(2(E-V(x))/m}}\bigg)^{-1}
\end{align}
The Hill-Wheeler formula for tunneling through a parabolic barrier is suitable here due to the proximity of the molecule energies in the tweezer to the top of the barrier, which makes the parabolic approximation to the barrier appropriate. The Hill-Wheeler transmission coefficient depends on the energy gap from the top of the parabolic barrier $E_{\text{gap}} = V_{max} - E$ and the curvature of the barrier $\omega_b = \sqrt{-V''(x_b)/m}$.
\begin{align}
     T_{HW} = (1 + e^{{\frac{2\pi E_{\text{gap}}}{\hbar \omega_b}}})^{-1}
\end{align}
To extract tunneling timescales for single molecule isolation fidelity, the gradient is chosen such that the lowest quasi-bound state energy $E$ satisfies $E_{\rm gap} = V_{\max} - E = \Delta E + 50$ Hz. This 50 Hz offset is included as it is small compared to the interaction shift and avoids numerical instability when estimating the tunneling timescales as would happen when $E_{\rm gap} \to 0$.

The tunneling timescale $\tau$ defined in Eq.~\eqref{eq:tautimescale} is the single molecule tunneling timescale evaluated at the relevant energy gap. Let $\tau^{(2)}$ ($\tau^{(1)}$) represent the tunneling timescale in the interaction-shifted (unshifted) configuration. Under the assumption of uncorrelated sequential one-body tunneling, the escape rates of the molecules add. Thus, we assign $\tau_A = \tau^{(2)}/2$ as the timescale for the first of the two molecules to escape from the doubly-occupied tweezer. The escape timescale for the single remaining molecule from the tweezer is assigned $\tau_B = \tau^{(1)}$.

\section{ Bound-state-free two-body loss rates \label{app:twobodyloss}}
\begin{figure}[ht]
    \centering
    \includegraphics[width=\linewidth]{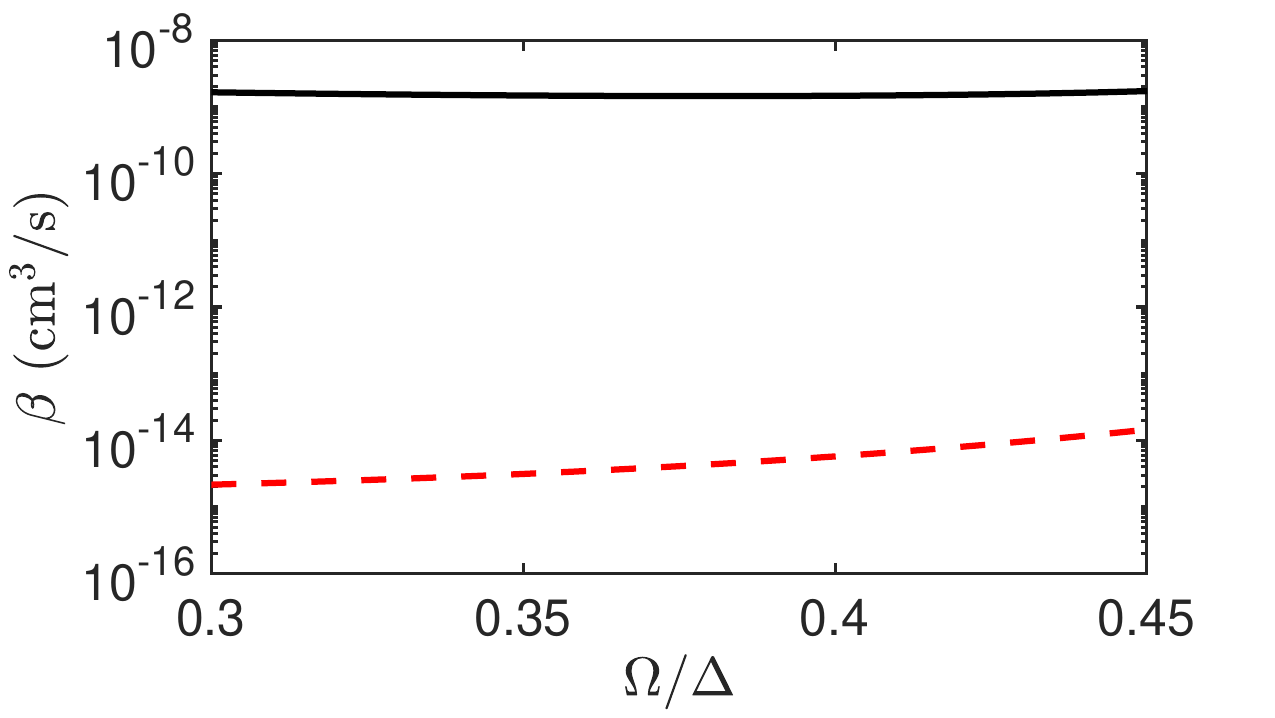}
    \caption{ Two-body rate coefficients as a function of Rabi frequency for a fixed detuning $\Delta = 1.3\Delta_F$. }
    \label{fig:rateCoefficient_vs_Omega_Delta1_3DeltaF}
\end{figure}

To estimate the two-body loss rate from scattering rate coefficients, we require an estimate of the density of molecules in the optical tweezer. Modeling an optical tweezer as a 3D harmonic oscillator with geometric mean trap frequency $\overline{\omega}$, the density at the origin given $N$ molecules of mass $m$ is estimated as
\begin{align}
    n_0 = \frac{N}{\pi^{3/2}} 
    \left(\frac{m\overline{\omega}}{\hbar}\right)^{3/2}. 
\end{align}
This assumes non-interacting molecules, which provides a conservative upper bound estimate for the density (see App.~\ref{app:Fermis_golden_lossrates}). The two-body loss rate coefficient $\beta_{\rm loss}$ is then used to produce the two-body loss timescale via
\begin{align}
    \tau_C = (\beta_{\rm loss} n_0)^{-1}.
\end{align} 
To eventually perform spilling, we in fact employ a ratio of $\Omega/\Delta = 0.3$ instead of 0.4 that minimizes the long-range well, as the former provides the largest in-tweezer interaction shift. 
As such, we compute the rate coefficient between these two ratios as plotted in Fig.~\ref{fig:rateCoefficient_vs_Omega_Delta1_3DeltaF}, showing that the loss rate in fact decreases toward $\Omega/\Delta = 0.3$.  

\section{ Probabilistic model for spilling fidelity \label{app:prob_model_description}}

As described in the main text, we compute the probability that there remains a single molecule in the trap after applying a field gradient for time $t$. While two molecules occupy the trap, two processes can occur: a single molecule can tunnel out (Event A, timescale $\tau_A$) or a two-body loss event can occur (Event C, timescale $\tau_C$). If a single molecule tunnels out at time $t'$, then the remaining molecule tunneling process can still occur (Event B, timescale $\tau_B$) during the interval from $t'$ to $t$.  We model each of these processes as an exponential decay. We determine the probability $P_1(t)$ of single molecule occupancy by integrating over all of the possible times for Event A to occur up to the spilling time $t$, requiring that two-body loss does not occur before time $t'$ and the remaining molecule does not tunnel out between time $t'$ and $t$.
\begin{align}
    P_1(t) = \int_{0}^{t} \frac{1}{\tau_A}e^{-\frac{t'}{\tau_A}}\cdot e^{-\frac{t-t'}{\tau_B}} \cdot e^{-\frac{t'}{\tau_C}}dt'
\end{align}
Defining $\Lambda = \frac{1}{\tau_A} + \frac{1}{\tau_C}$, performing the integral above, and finally optimizing $P_1(t)$, we determine the optimal spilling fidelity with the required optimal spilling time.
\begin{align}
    P_{\text{opt}} = \frac{1}{\Lambda \tau_A}(\Lambda \tau_B)^{\frac{1}{1 - \Lambda \tau_B}},\qquad t_{\text{opt}} =\frac{\log(\Lambda \tau _B)}{\Lambda - \frac{1}{\tau_B}}
\end{align}
This calculation demonstrates the importance of (1) separating the one- and two-molecule tunneling timescales and (2) minimizing the two-body loss rate.

\section{Microwave uniformity across the array\label{app:microwave_uniformity}}

In experimental apparatuses, reflections can interfere with the incident microwave field and reduce its spatial uniformity. To produce an apparatus-independent estimate of this wavelength-scale variation, we consider the case in which the incident microwave interferes with a perfect co-polarized reflection, producing a standing wave. The Rabi frequency therefore takes the form:
\begin{align}
    |\Omega(z)| = 2\Omega_0|\cos(kz + \phi/2)|
\end{align}
where $k = 2\pi/\lambda$, and by setting $\phi=0$, the antinode of the standing wave sits at the center of the array. The ac-dc(X) shielding scheme addresses the dressed rotational state transition $|\tilde{N} = 1, m_N = 0\rangle \to |\tilde{N} = 2, m_N = \pm   1\rangle$ at the field $\rm E_{dc} = 2.366$ kV/cm. This transition frequency is $f = 6.2186$ GHz, equivalent to a wavelength of $\lambda = 4.82$ cm. 

With $L$ denoting the full extent of the array along the microwave propagation direction such that $z \in [-L/2, L/2]$, the fractional change of the Rabi frequency is computed by noting that the maximum occurs at $z = 0$ and the minimum will occur at the edge of the array. 
\begin{align}
    \frac{\delta \Omega}{\Omega} \equiv \frac{\Omega_{max} - \Omega_{min}}{\Omega_{max}} = 1 - \cos\bigg(\frac{\pi L}{\lambda}\bigg) \approx \frac{\pi^2}{2}\bigg(\frac{L}{\lambda}\bigg)^2
\end{align}
For a representative array size of $L = 200 \mu$m, the fractional change $\frac{\delta \Omega}{\Omega}  \approx 8.49 \cdot 10^{-5}$.

\section{ Estimating {loss rates} for trapped molecules \label{app:Fermis_golden_lossrates} }

Apart from scattering calculations, Fermi's golden rule also provides an estimate of the two-body loss rate \cite{Napolitano94_PRL, Julienne97_JRNIST} of two tweezer ground state molecules:
\begin{align} \label{eq:FGRlossrate}
    \Gamma_{\rm loss}
    &=
    \frac{ 2\pi }{ \hbar }
    \int \frac{ d^2\hat{\boldsymbol{k}} }{ 4\pi }
    \left| \langle{\boldsymbol{k}_{o}}| 
    V_{c} 
    |{\rm gs}\rangle \right|^2,
\end{align}
where $| \boldsymbol{k}_{o} \rangle$ is the energy-normalized outbound state with energy associated with the inelastic threshold, 
$V_{c}$ is the coupling matrix element between the elastic and energetically closest inelastic thresholds, and $|{\rm gs}\rangle$ is the trapped two-molecule ground state wavefunction obtained from imaginary time evolution.    
This estimate assumes that the inelastic loss rate is far greater than the loss rate due to molecular tunneling into the short-range, the latter of which is typically orders of magnitude smaller in NaCs \cite{Karman25_PRXQ, Wang26_arxiv}. 
The inelastic thresholds are typically $\sim 10$ MHz below the prepared molecular state while the trap depth is $< 100$ kHz, resulting in a large gain in kinetic energy of the lost molecules that escape essentially as free particles. 
As a result, the outbound wavefunction is well approximated by the energy-normalized plane wave
\begin{align}
    \langle \boldsymbol{r} | \boldsymbol{k}_{o} \rangle
    &= 
    \frac{ \sqrt{ \mu k_{o}(\boldsymbol{r}) } }{ \hbar  }
    \frac{ e^{ i \boldsymbol{k}_{o}(\boldsymbol{r}) \cdot \boldsymbol{r} } }{ ( 2 \pi )^{3/2}  },
\end{align}
with wavenumber $k_{o}(\boldsymbol{r}) = \sqrt{ 2 \mu [ E_{\rm gs} - \varepsilon_{o} - V_{o}(\boldsymbol{r}) ] / \hbar^2 }$ where $E_{\rm gs}$ is the relative ground state energy, $\varepsilon_{o}$ is the energy of outbound threshold relative to the incident threshold, and $V_{o}(\boldsymbol{r})$ is the potential energy surface of the diabatic inelastic channel.

The matrix element is evaluated as the discretized numerical integral
\begin{align}
    \langle \boldsymbol{k}_{o} | T |{\rm gs}\rangle
    &=
    \int_V 
    d^3\boldsymbol{r}
    \frac{ \sqrt{ \mu k_{o}(\boldsymbol{r}) } e^{ i \boldsymbol{k}_{o}(\boldsymbol{r}) \cdot \boldsymbol{r} } }{ (2\pi)^{3/2} \hbar } 
    V_{c}(\boldsymbol{r})
    \psi_0(\boldsymbol{r}),
\end{align}
where $V$ is the volume over which we perform numerical integration, and $\int_V d^3\boldsymbol{r} | \psi_0(\boldsymbol{r}) |^2 = 1$.  
Unfortunately, the three-dimensional spatial grid used for ground state search utilizes discrete cells with widths much larger than $2\pi / k_{o}$. 
To get around this issue, we assume that $V_c(\boldsymbol{r})$ and $V_o(\boldsymbol{r})$ are constant over each grid cell, so that the integral over each grid cell is approximated by 
\begin{align}
    & \int_{\rm cell}
    d^3\boldsymbol{r}'
    \frac{ \sqrt{ \mu k_{o}(\boldsymbol{r}) } e^{ i \boldsymbol{k}_{o}(\boldsymbol{r}) \cdot \boldsymbol{r}' } }{ (2\pi)^{3/2} \hbar }
    V_{c}(\boldsymbol{r})
    \\
    \approx &\:
    \left. 
    V_{c}(\boldsymbol{r})
    \frac{ \sqrt{ \mu k_{o}(\boldsymbol{r}) } }{ (2\pi)^{3/2} \hbar }
    \right|_{\boldsymbol{r} = \boldsymbol{r}_{\rm cell}} 
    \int_{\rm cell}
    d^3\boldsymbol{r}'
    e^{ i \boldsymbol{k}_{o}(\boldsymbol{r}_{\rm cell}) \cdot \boldsymbol{r}' }, \nonumber
\end{align}
where $\boldsymbol{r}_{\rm cell}$ is the centroid of the cell. The integral within each cell is then
\begin{align}
    \int_{\rm cell} d^3\boldsymbol{r}'
    e^{ i \boldsymbol{k}_{o} \cdot \boldsymbol{r}' } 
    &=
    \int_{V_{\rm cell}} d^3\boldsymbol{r}'
    e^{ i \boldsymbol{k}_{o} \cdot (\boldsymbol{r}' - \boldsymbol{r}_{\rm cell}) } \\
    &=
    e^{ -i \boldsymbol{k}_{o} \cdot \boldsymbol{r}_{\rm cell} }
    \int_{V_{\rm cell}} d^3\boldsymbol{r}'
    e^{ i \boldsymbol{k}_{o} \cdot \boldsymbol{r}' } \nonumber\\
    &=
    \frac{ 8 }{ k_{o}^3 }
    \frac{ \prod_i \sin[ ( \hat{\boldsymbol{k}}\cdot\hat{\boldsymbol{r}}_i ) {k}_{o} \Delta{r}_i / 2 ] }{ \prod_i |\hat{\boldsymbol{r}}_i| }
    e^{ -i \boldsymbol{k}_{o} \cdot \boldsymbol{r}_{\rm cell} }, \nonumber
\end{align}
where $\boldsymbol{r}_{\rm cell}$ is the centroid of the cell and $V_{\rm cell}$ is its volume with edge lengths $\Delta{r}_i$,
from which the integral over the entire simulation volume is then given by the sum over all cell terms.
Finally, the integral over $\hat{\boldsymbol{k}}$ is taken using Gauss-Legendre quadratures and Simpson's integration over the inclination and azimuthal angles respectively.   
We neglect complications of the short-range by ignoring the region within $r < 700 a_0$ in our calculations. 

\begin{figure}[ht]
    \centering
    \includegraphics[width=\linewidth]{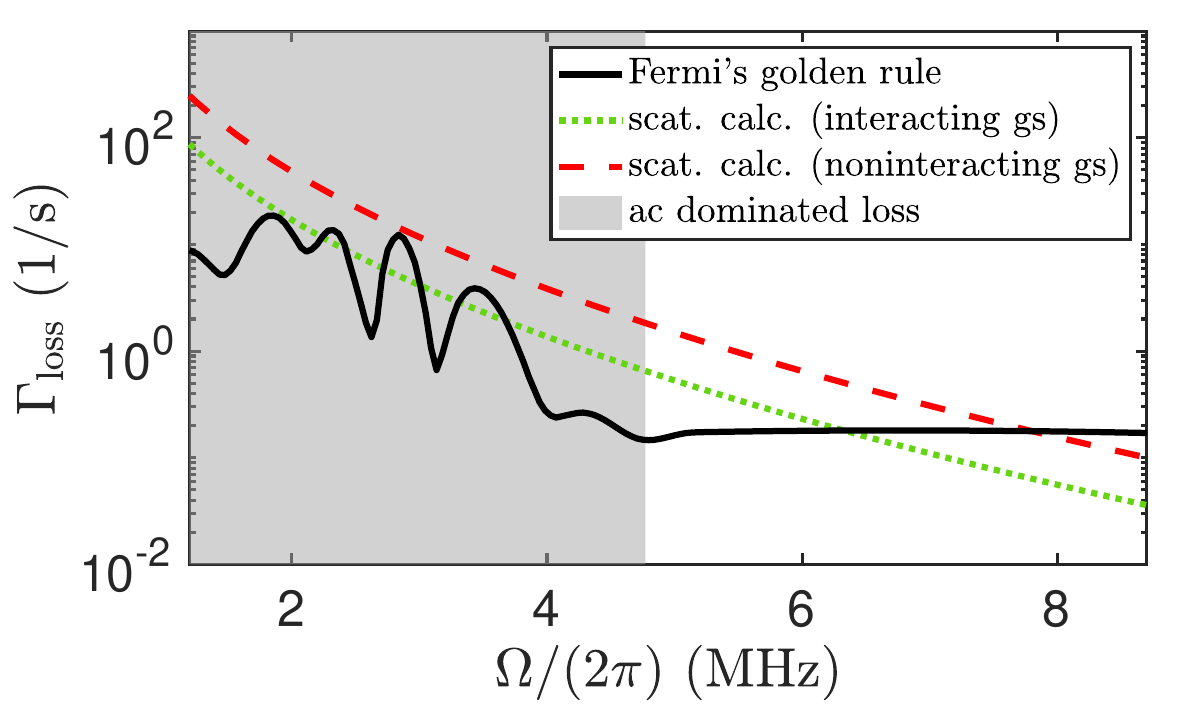}
    \caption{ Two-body molecular loss rate as a function of Rabi frequency with $\Omega = 0.4 \Delta$, obtained using Fermi's golden rule (solid black curve), compared to scattering calculations with the noninteracting ground state density (dashed red curve), and the interacting ground state density (dotted green curve). The tweezer potential is assumed harmonic with frequency $(\omega_x, \omega_y, \omega_z) = 2\pi \times (1, 5, 5)$ kHz. The gray shaded region is dominated by microwave-induced inelastic loss, while the unshaded region is dominated by loss to the dc F\"orster channel. }
    \label{fig:FGRlossrate_vs_Omega}
\end{figure}

The calculated loss rates are plotted in Fig.~\ref{fig:FGRlossrate_vs_Omega} (solid black curve), having fixed the ratio $\Omega/\Delta = 0.4$.  
These loss rates feature large variations for lower values of $\Omega$ compared to those obtained from scattering calculations (see Fig.~\ref{fig:rateCoefficient_vs_Omega}). 
Varying $\Omega$, and correspondingly $\Delta$, shifts the Condon point between the initial and closest inelastic diabatic channels, causing the observed sharp dips in $\Gamma_{\rm loss}$ similar to photoabsorption interference resonances \cite{Fano68_RMP}. 
At large Rabi frequencies of $\Omega/(2\pi) > 4.8$ MHz, the loss rate becomes nearly constant at $\Gamma_{\rm loss} \approx 0.17$ s$^{-1}$ when the closest inelastic threshold is that induced by the dc field $|F\rangle$, with a detuning from the elastic threshold of $\hbar\Delta_F + \varepsilon_+$ that weakly varies with $\Omega$.  

Loss rates are obtained from the scattering calculations of $\beta_{\rm loss}$ through multiplication by the two-molecule ground state density.  
This ground state is taken either as the ground state of two harmonically trapped noninteracting molecules (dashed red curve in Fig.~\ref{fig:FGRlossrate_vs_Omega}), or the interacting ground state from imaginary time evolution (dotted green curve in Fig.~\ref{fig:FGRlossrate_vs_Omega}).  
The latter is always lower than the former due to the large shielding core, and is computed from the numerical ground state in relative coordinates through
\begin{align}
    n_{0}(\boldsymbol{r}_1)
    &=
    \int d^3\boldsymbol{r}'_1 d^3\boldsymbol{r}'_2
    \delta^3(\boldsymbol{r}'_1 - \boldsymbol{r}_1)
    |\Phi(\boldsymbol{r}'_1, \boldsymbol{r}'_2)|^2 \nonumber\\
    &=
    \int d^3\boldsymbol{r}' d^3\boldsymbol{R}'
    \delta^3(\boldsymbol{R}' + \boldsymbol{r}'/2 - \boldsymbol{r}_1)
    |\Psi_0(\boldsymbol{R}')|^2
    |\psi_0(\boldsymbol{r}')|^2 \nonumber\\
    &=
    \int d^3\boldsymbol{r}' 
    |\Psi_0(\boldsymbol{r}_1 - \boldsymbol{r}'/2)|^2
    |\psi_0(\boldsymbol{r}')|^2,
\end{align}
where $\Psi_0(\boldsymbol{R}) = [M \overline{\omega} / (\pi\hbar)]^{3/4} \exp[ -M \sum_i \omega_i R_i^2 / (2\hbar) ]$ is the center-of-mass trapped ground state, while $\boldsymbol{r} = \boldsymbol{r}_1 - \boldsymbol{r}_2$ and $\boldsymbol{R} = (\boldsymbol{r}_1 + \boldsymbol{r}_2) / 2$ are the relative and center-of-mass coordinates respectively. 
The density for two molecules is then $2 n_0(\boldsymbol{r})$.  
When the noninteracting ground state is used, the loss rate from scattering calculations is far greater than from Fermi's golden rule at $\Delta = \Delta_F$. The loss rates are similar when the interacting ground state is used instead.

\nocite{*}
\bibliography{main.bib}

\end{document}